\documentclass[11pt]{article}
\usepackage{amsmath,epsfig}

\newcommand{\Ds}{\text{$/$\hspace{-0.37cm} $\nabla$}}
\newcommand{\ps}{\text{$/$\hspace{-0.37cm} $\partial$}}
\newcommand{\D}{$\overline{\rm D}$}

\newcommand{\Z}{{\bf Z}}
\newcommand{\Tr}{{\rm Tr\,}}
\newcommand{\STr}{{\rm STr\,}}
\newcommand{\diag}{{\rm diag}}
\renewcommand{\Im}{{\rm Im\,}}
\renewcommand{\Re}{{\rm Re\,}}

\newcommand{\1}{{\bf 1}}

\begin{document}

\title{\vbox{
\baselineskip 14pt
\hfill \hbox{\normalsize
} \\
\hfill \hbox{\normalsize {SNUTP-06-006}} } \vskip 2cm  Intersecting
Brane World \\ From Type I Compactification}

\author{{\large Kang-Sin Choi}
\\ \\
%\address
\normalsize \it Physikalisches Institut, Universit\"at Bonn,
Nussallee
 12, D-53115 Bonn, Germany and \\ \normalsize \it
 Center for Theoretical Physics, Seoul National
  University, Seoul 151-747, Korea \\
{\tt kschoi@th.physik.uni-bonn.de }
 }

%\begin{frontmatter}
\date{}
\maketitle
\thispagestyle{empty}

\begin{abstract}
We elaborate that general intersecting brane models on
orbifolds are obtained from type I string compactifications and
their $T$-duals. Symmetry breaking and restoration occur via
recombination and parallel separation of branes, preserving
supersymmetry. The Ramond--Ramond tadpole cancelation and the toron
quantization constrain the spectrum as a branching of the adjoints
of $SO(32)$, up to orbifold projections. Since the recombination
changes the gauge coupling, the single gauge coupling of type I
could give rise to different coupling below the unification scale.
This is due to the nonlocal properties of the Dirac--Born--Infeld
action. The weak mixing angle $\sin^2 \theta_W=3/8$ is naturally
explained by embedding the quantum numbers to those of $SO(10)$. \\ \\
Keywords: unification, intersecting branes, recombination, weak
mixing angle.
\end{abstract}

\maketitle

\section{Introduction}

There are two nice descriptions of gauge theory  from string theory,
which we expect, in the end, to bring forth the Minimal
Supersymmetric Standard Model (MSSM). One is a closed string on which
the charge is distributed homogeneously and gives rise to the current
algebra. This idea is realized by heterotic string, and by
compactification on a suitable manifold, we obtain semi-realistic
theories close to MSSM. The other is open string description of type
I and type II strings. There the gauge degree of freedom, so called
Chan--Paton factor, is assigned on the endpoints of open strings, whose
modern understanding is provided by D-brane stacks, in the $T$-dual
picture. In particular, the chiral fermions emerge on intersections
of branes at angle \cite{BDL}. This provides the framework of a
collection of models called intersecting brane world \cite{BCLSU}.

The most unsatisfactory point of the brane picture has been the lack
of understanding on unification, thus conventional models have
resorted to bottom-up approach. However, if we consider string
theory as the first principle from which we reduce our world, there
must be {\em unification}. Then the theory is consistent by
construction. For instance, anomaly cancelation is automatically
satisfied if we spontaneously break an unified theory, which is
anomaly free by nature. Also from the bottom, the running gauge
couplings from MSSM and unification around $10^{16}$ GeV \cite{GQW}
is quite appealing. For the resulting low-energy vacua to be
explained by spontaneous symmetry breaking, we should have an
unified gauge group and coupling at the symmetry breaking scale. In
this paper, we consider this possibility.

The Higgs mechanism via adjoint representation is easy to
understand. The nonabelian gauge group is realized by a stack of
coincident branes. Their positions are denoted by matrix-valued
vector $X^m$ also transforms as an adjoint. When they are diagonal
in group space, the geometric information on D-branes translates to that
on the gauge setup. There it is a Wilson line
\begin{equation} \label{adjHiggs}
 A_m = {1 \over 2 \pi \alpha'} X^m = \begin{pmatrix} a_1 & & & \\ & a_2 & & \\ & & \ddots &
 \\ & & & a_n \end{pmatrix},
\end{equation}
where $1/(2 \pi \alpha')$ is the string tension. In general they
represent a broken phase of $U(1)^n \subset U(n)$. When some
elements are same, or some branes are coincident, the gauge symmetry
is enhanced to $\prod U(n^{i})$, where $n^{i}$ is the number of
the same elements. We can see that the symmetry breaking corresponds
to {\em brane separation}. The parallel branes are
half-Bogomoln'yi-Prasad-Sommerfield (BPS) system, thus parallel
separation does not break further supersymmetries. Also the
D-flatness is not spoiled since D-term potential is given by
\begin{equation} \label{dtermpot}
 \Tr [X^m,X^n]^2
\end{equation}
from dualizing the gauge kinetic term.

However to explain the chiral nature of matter fermions, we need
intersecting branes. Still in the intersecting case, the parallel
separation moduli survive \cite{BLS,CLLL}. However we have also a
deformation between intersecting and parallel branes, known as
{\em recombination} \cite{Choi06,HT,CIM02,EGHK,DZ,KW,CK}. In the
non-supersymmetric case, the instability is recovered by
recombination, setteling down to minimal volume setup. However we see
that if protected by supersymmetry, the recombination costs no
energy.

In this paper we consider compactifications on orbifold with suitable
orientifold planes.
To have supersymmetric models, we require some ``F-flatness'' condition, which
is a generalization of self-dual condition \cite{MMMS}. Also we have
the Ramond--Ramond (RR) charge conservation, realized as tadpole
cancelation condition, which guarantees the absence of
anomalies. We will see that every configuration
satisfying these have the same energy, in particular, to type I
compactification on the same orbifold. Thus we naturally expect that
also deformation along the flat directions could explain the
spontaneous symmetry breaking of gauge group from {\em type I
compactification} \cite{Choi06}.
During the recombination, although the intersection points, hence
local chiralities, are changed, but the charge is conserved thus the
total amount of anomalies is not affected. These points are examined in
section 3. Such unification picture is also natural from the charge
embedding, as to be discussed in section 2 in the modern D-brane picture.
From quantization condition for tilted branes, charges are
embedded into adjoint representations, which then belong to those of
$SO(32)$ and a bifundamental
representations occur as off-diagonal entries of them,
like the $X,Y$ bosons in Georgi--Glashow unification \cite{GG}, for
instance.  In section 4, we analyze the
gauge coupling unification using fluctuation spectrum analysis of
Dirac--Born--Infeld (DBI) action \cite{Choi06,AHT,DST}. Also we
comment on the desirable weak mixing angle $\sin^2 \theta = 3/8$ at
the unification scale.

\section{Adjoint embedding}

We begin with the proper description of intersecting branes by
embedding representations into an adjoint of nonabelian group.
Bifundamental representations can arise from branching of adjoints.
They are the only possible representation from open
strings\footnote{In heterotic string, a twisted sector state can be
an antisymmetric tensor representation, from twisted algebra, which
mixes higher grade state \cite{Choi04}.}.

%\subsection{Separation of branes} \label{subsec:separation}

\subsection{Toron quantization}

Toron is a quantum of constant magnetic field on the torus
\cite{tH}.\footnote{A clear exposition, in both the gauge theory and
  D-brane points of view, can be found in refs.
\cite{vB84,GR,Ra,CIM,Ta}.} They are realized by various D-branes and
their bound states.

\subsubsection*{Two dimensional case}

Consider a D1-brane wrapping on a two torus $T^2$ in 12-directions.
In string theory a torus $T^2$ as a target space is specified by two
complex numbers, namely the complex structure $\tau$ and K\"ahler
modulus $B + i A$ where $A = L^2 \Im \tau, L \in {\bf R}, \tau \in
{\bf C}$ and $B$ is the NSNS antisymmetric tensor. It is equivalent
to choice of basis vectors $e_1 = L ,e_2 = L \tau$ and a NSNS field
$B$. This D1-brane is specified by two integral numbers $(n,m)$,
namely straightly winding $n,m$ times along $e_1,e_2$ directions
respectively. We easily see that the cycle is of the minimal length
for given $(n,m)$ and described as
\begin{equation} \label{growvev}
  \frac 1 {2\pi \alpha'} X^2 =  F_{12} X^1 +  A^0_{1},
\end{equation}
where
\begin{equation} \begin{split} \label{qtz}
 F_{12}&={2\pi \over A} {m \over n} \1_n \equiv \tan
 (g \theta) {2 \pi \over  A} \1_n \\
 A^0_{1} &= {2 \pi \over A}\frac1
 n \diag(0,1,\dots,n-1) + c,
\end{split} \end{equation}
with the string coupling $g$, the identity matrix of rank $n$ and a
constant $c$. $\theta$ is interpreted as the relative angle to
1-direction. Although we have a single cycle, we employed matrix
structure in the group space in the spirit of (\ref{adjHiggs}). We
regard it as a {\em single} brane identified inside the torus,
rather than $n$ slices of equally separated D-brane with permutation
symmetry. This quantization arises by requiring {\em closedness} of
the brane, since it can only end on other D or NS branes, not on the
empty space. Although a homologous setup with $n$ parallel D1 branes
along 1-direction and $m$ branes along 2-direction has the same
charge, it is a different setup.

The gauge degree of freedom a fluctuation around this flux
(\ref{growvev}). Although {\em embedded} in $U(n)$, the actual
degree of freedom is smaller $U(p)$ for $p \equiv \gcd (n,m)$. We
can check that $p$ is the maximal effective number of coincident
branes, easily seen from the geometric picture. In the gauge theory
point of view, there can be at most $p$ permutation symmetry $S_p$
in (\ref{growvev}). It also means,  having less permutation
symmetry, we can break the above $U(p)$ gauge group by parallel
separation, adjusting constant matrix in (\ref{qtz}) in analogy with
the Wilson line case (\ref{adjHiggs}).

$T$-duality in 2-direction maps the two to D0 brane and D2 brane.
Two moduli are exchanged and $L' = 2 \pi \alpha'/L$. The above becomes
\begin{equation} \label{d0d2flux}
  A_2 = F_{12} X^1, \quad A_1 = 0,
\end{equation}
which is a form of gauge fixed constant magnetic flux.
From the rank
of the matrix $n$ slices of D2 brane describe worldvolume $U(n)$
gauge theory. This is the consistent solution satisfying the 't
Hooft boundary condition \cite{tH,AHT}. The condition (\ref{qtz})
become quantization condition.
\begin{equation} \label{toronqtz}
 c_1 = \frac1{2 \pi} \int d^2 x \Tr F_{12} = m,
\end{equation}
with $m$ being integer, indeed, from vortex quantization. From the
Chern--Simons coupling, the constant flux is interpreted as 0-brane
charge \ $i\mu_2 2 \pi \alpha' \int_{T^2 \times M} \Tr  F_{12} \wedge
C_0 = i m \mu_0 \int_M C_0 $ so we have $m$ D0-branes.
This is interpreted as D0-D2 bound state and in
fact D0 brane has melt into D2 brane, leaving the flux
(\ref{d0d2flux}) on it. This bound state is again 1/2 BPS state
\cite{GGBL}.

By a series of $SL(2,\Z)$-dual there is a configuration where we have
$p$ zero branes only on $T^2$ with the moduli space $(T^2)^p/S_p$
\cite{GR}, thus we again see that the resulting gauge group is $U(p)$.

\begin{figure}[t]
\begin{center}
\includegraphics[height=3.1cm]{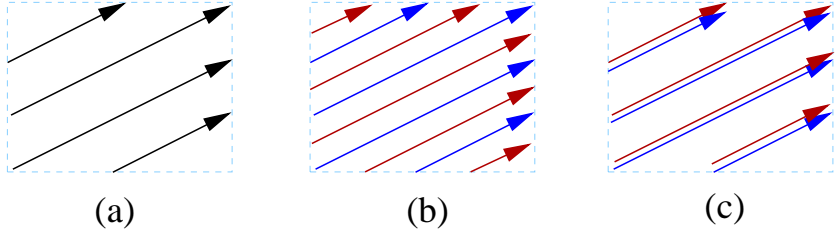}
\end{center}
\caption{A cycle with winding number $(3,2)$ describes gauge
  group $U(1)$, as a broken phase of $U(3)$ due to the quantization condition
  (a). A $(6,4)$ cycle, a double copy of (3,2), generically has $U(1)\times U(1)
 \subset U(6)$ (b), and can be enhanced to $U(2) \subset U(6)$
 when branes are coincident (c).}
\end{figure}

Note that $T$-dual in 12-direction, we exchange $n$ and $m$, i.e. we
have $U(m)$ gauge symmetry with the first Chern number $n$. This
symmetry is not visible in gauge theory since it only take into
account K-theory charge, but natural in the sense it merely
corresponds to exchanging coordinates $x^1$ and $x^2$. (In higher
dimensional case, in the shrinking instanton limit, we see both
gauge group.) Thus even if we have $U(1)$ gauge symmetry, the theory
is naturally embedded in $U(n)$ or $U(m)$ gauge theory. In other
words, the natural embedding is {\em nonabelian}.

A pair of parallel D1-branes can make further {\em bound state}
between themselves, which is equivalent to {\em one D1-brane winding
twice} the same cycle. The $SU(2)$ degree of freedom decouples and
the low energy degree of freedom is $U(1)$ gauge theory \cite{Sen1}.
For the above $(n,m)$ cycle, we see there is $p$ kinds of bound
states. This bound state is not distinguishable, by the charge, or
Chern number.

When we have more than one stacks in the same torus, we may
generally have a number of cycles. The resulting toron background is
expressed as
\begin{equation}
 F_{12} = {2\pi \over A}
 \begin{pmatrix} {m^1 \over n^1} \1_{n^1} & & \\
 & \ddots & \\
 & &  {m^k \over n^k} \1_{n^k} \end{pmatrix}.
\end{equation}
Without loss of generality, we can make the setup that each block
satisfies the above condition (\ref{toronqtz}).  However
the important measure is the first Chern number $c_1 = \sum_1^k
m^{a}$. It seems natural to embed the field strengths of all the
stacks into a single nonabelian one. Evidently, by the form of
matrix, it is a Higgs phase of
\begin{equation}
  U(n^1) \times \cdots \times U(n^k) \subset U(n), \quad  n =
\sum_{a=1}^k n^{a},
\end{equation}
generalizing (\ref{adjHiggs}). Although this setup cannot directly
come from the single stack (\ref{qtz}), all of them can be regarded to
reside specific points in moduli space of $U(n)$ gauge theory.
This is clear in the bound state picture that we have
$n$ D2- and $c_1$ D0-branes forming $k$ bunch of bound states. How
differently they can bound determines the setup.
Shortly we will see that there is an upper limit on $m$ and $n$ from
the tadpole cancelation condition.

For general compactifications on $\Z_N$ or $\Z_M \times \Z_N$ orbifolds
the unit torus is not rectangular but parallelogram with $\tau \ne 1$,
 to be compatible with orbifold group. This tilting is also required
 even if we can have rectangular tori, since intersection number
accounting for the number of generations, is even in general,
compared to the observed number, three. In a the case of our
interest, worldsheet parity projection $\Omega$ projects out
antisymmetric NSNS $B$ field. When we compactify it, $B$ field in
the compact dimension can assume discretized, half-integral values
$0,\frac12$ modulo 1 \cite{AB,BGK}. In the $T$-dual picture, this is
the compatibility of orientifold action $\Omega R$ on torus, and
$\Re \tau=0,\frac12$ modulo 1. In the half-integral case, the RR
charge of orientifold is reduced by half, which leads rank and
multiplicity reduction. For the $\Re \tau =\frac12$ case, it is
convenient to define
$$ \tilde m = m + \frac12 n.$$
In the literature, they are called A- and B-torus respectively.
The angle condition is related to
\begin{equation} \label{anglecond}
 \tan g \theta = {m + n \Re  \tau \over n \Im \tau
 } = \left\{ \begin{matrix}
\displaystyle{  m \over n\Im \tau} & (\Re \tau =0)   \\ \\
 \displaystyle {  \tilde m \over n\Im \tau} & (\Re \tau =\frac12)
 \end{matrix} \right. .
\end{equation}
Evidently, for the case of titled torus, the discussion is not
changed if we replace a cycle number with a tilted one.

\subsubsection*{Four and higher dimensions}

We can generalize the above idea to arbitrary dimensions. There are
six two-cycles in four-torus, i.e. dim$\, H_2(T^4,\Z) = 6$. They can
also be represented by six integral numbers. In $T$-dual space, the
system looks as D0-D4 bound states and the six numbers are related
to Chern numbers
\begin{equation}
(N,c_1^{12},c_1^{34},c_2,c^{14}_1,c^{23}_1).
\end{equation}
The $c_i$ denotes the $i$th Chern numbers in $T$-dual picture, where
we have D0 and D4 branes and the superscript denotes the direction.
These numbers can be reshuffled by $T$-dualities $SL(4,\Z)$, and
with it we may make three of them to be zero \cite{vB84,GR97}. For
convenience we will consider the factorizable cycle $Nc_2= c_1^{12}
c_1^{34},c_1^{14}=0$, which is understood as that of direct product
of two-torus. However, the discussion will not be changed for the
non-factorizable cycle. The cycles are denoted by three independent
numbers
\begin{equation} \label{t2cherns}
 (n_1,m_1)(n_2,m_2) = (n_1 n_2, n_1 m_2, m_1 n_2, m_1 m_2) =
 (N,c_1^{14},c_1^{23},c_2).
\end{equation}
Therefore we have
\begin{equation} \label{4quant}
 F_{12} = {2\pi \over A_1} \begin{pmatrix} {m^1_1 \over n^1_1}
  \1_{N^1} & & \\
  & \ddots & \\
 & & {m^k_1 \over n^k_1}  \1_{N^k } \end{pmatrix}, \quad
 F_{34} = {2\pi \over A_2} \begin{pmatrix} {m^1_2 \over n^1_2}
  \1_{N^1 } & & \\
  & \ddots & \\
 & & {m^k_2 \over n^k_2}  \1_{N^k} \end{pmatrix}.
\end{equation}
Beside the multicomponent in the different directions, only the
slight different is that the rank of each unit matrix is now $N^{a}
= n_1^{a} n_2^{a}$. The slopes stay in the same form, but now they
can be also expressed by Chern numbers $m/n = c^{ij}_1/ c_2$. Again,
we see that the system is naturally embedded into $U(N)$ theory,
where $N=\sum N^{a}$. Note that the quantization condition requires
the dimensions of diagonal blocks to be the same $I_{N^{a}}$,
although the slopes can be different. Each describes one stack of
branes. Therefore, as in the parallel brane case, it seems natural
and suggestive to embed all the branes into {\em one representation}
and regard the subalgebra as a {\em broken phase}. We will see the
upper limit of the rank will be given by tadpole cancelation
condition.

For the six dimensional case, the factorizable three cycle is represented as
\begin{equation} \label{18cycle} \begin{split}
 & (n_1,m_1)(n_2,m_2)(n_3,m_3) \\ & = (n_1 n_2 n_3, n_1 m_2 m_3, m_1 n_2
  m_3, m_1 m_2 n_3, m_1 m_2 m_3, m_1 n_2 n_3, n_1 m_2 n_3, n_1 n_2
  m_3) \\
 & = (N,c^{12}_2,c^{34}_2,c^{56}_2,c_3,c_1^{12},c_1^{34},c_1^{56})
\end{split} \end{equation}

We can straightforwardly generalize such toron descriptions to an
arbitrary number of extra dimensions. In such descriptions, all the
information is contained in the Chern numbers. As pointed out
\cite{Wi98}, these information can only take into account of
K-theory charge, not the cohomological information. So brane and
antibrane is not distinguishable, but only the charge difference
does. In this sense, the description as (\ref{toyBG}) is far from
complete. It cannot describe a brane lying vertically on the torus
$n_a = 0$ nor antibrane cycle $n_a<0$ and, more severely, when the
off-diagonal component of (\ref{toyBG}) is turned on, the
corresponding cycle is not same as $(n_a,m_a)$ any more, resulting
in different magnetic flux. Measuring D-brane charge by cohomology
rather than K-theory amounts to measuring a K-theory class by Chern
classes.

\subsection{Bifundamental representations}

One noticed that, for example, the $X,Y$ gauge
bosons in the conventional Georgi--Glasow $SU(5)$ unification
\cite{GG} have the correct quantum numbers, i.e. colors and weak
isospins $\bf (3,2)$,
 as (s)quarks. Of course
they do not have the desired $U(1)$ hypercharges, but provided that
they are also charged under additional $U(1)$ symmetries, after
diagonal symmetry breaking they may have the correct quantum
numbers. We see that such structure can explain the bifundamental
representation of MSSM matters embedded into adjoint
representations, which is the quantum number of a string stretched
between branes localized at an intersection.

Consider a symmetry breaking $G = U(p^a + p^b) \to H = U(p^a) \times
U(p^b)$. The case with more $U(n)$ for $H$ is to be
straightforwardly generalized by iteration. With a gauge choice we
can take
\begin{equation} \label{bifund}
 A_m =  {2\pi \over A} \begin{pmatrix}
  \ddots & & & & \\
  & \displaystyle {m^{a} \over n^{a}} \1_{N^{a}}  & \\
  & & \ddots & \\
  & & & \displaystyle {m^{b} \over n^{b}} \1_{N^{b}} & \\
  & & & & \ddots \end{pmatrix} X^n + A_m^0.
\end{equation}
Naturally $p^i=\gcd(n^{i},m^{i})$. In $4k+2$ dimension, the Dirac
spinor can be decomposed into two Weyl representations
\begin{equation} \label{DiracOp}
  \Ds =  \begin{pmatrix} 0 & i \ps + A_z \\
 i \bar \ps + A_{\bar z} & 0 \end{pmatrix}, \quad
  \Psi (z,\bar z) = \begin{pmatrix} \Psi_+ \\ \Psi_-
  \end{pmatrix}.
\end{equation}
Note that this kind of decomposition is possible when we have
$4k+2$ extra dimensions. Under the above gauge group $H$, an
adjoint fermion is naturally decomposed as
\begin{equation} \label{bifurep}
 \quad \Psi_\pm = \begin{pmatrix} \ddots & & & & \\
 & A_\pm & & B_\pm \\  & & \ddots & & \\  & C_\pm & & D_\pm & \\ & & & & \ddots \end{pmatrix}
\end{equation}
in the group space. Now solving Dirac equation, we find the gauge
transformation property of $B$ and $C$ are exactly bifundamentals
$(p^a,p^b), (p^a,\overline p^b)$, respectively. The solution is
given by biperiodic function, the Jacobi theta function. A detailed
solution is discussed \cite{CIM02}. The theory is {\em chiral}
\cite{CIM,vB82,AW}. This means, for example either quark $q$ or
antiquark $q^{\rm c}$ in (\ref{su7adj}) will be exclusively
massless. This can be seen by index theorem. It is well known that
in the nontrivial background of solitons and instantons, the Dirac
operator (\ref{DiracOp}) has {\em chiral} zero modes. The Dirac
operator is decomposed as $\Ds_{4k+6} = \Ds_4 + \Ds_{4k+2}$ and, for
$4k+2$ extra dimensions, the chirality is correlated $\Gamma^{4k+3}
= \Gamma^5$. The difference of the number of left and right mover
zero modes, contributed from the $a$ and $b$ block, is the Chern
number,
\begin{equation} \label{diracidx}
 I_{ab}  =
 \text{index}_{Q,ab} \Ds_6 = {1 \over 3!(2 \pi)^3} \int \Tr_{Q,ab}
F^3
\end{equation}
where the trace is over gauge charge $Q$ of commutant group $L$ to
$H$, completely determined by branching and is on the $a$th and $b$th
block. Plugging (\ref{bifund}),
we see that the each quark and lepton is chiral. Moreover, there
are $I_{ab}$ degenerate bifundamental solutions for each
off-diagonal solution. One can confirm that this is the same as
intersection number
\begin{equation} \label{intersecnum}
  I_{ab} = \prod_{i=1}^3 ( m^a_i n^b_i - n^a_i m^b_i)
\end{equation}
of D-branes, which can explain the number of generations. It is
determined by the topology of extra dimensions. Note that gauge
bosons of $H$ are always massless, nonchiral and non-degenerate. As
the expression indicates, in general there are massless chiral
fermions, whereas the scalar partners are massless only when there
is a supersymmetry.

A supporting evidence of this picture is suggested in the dual {\it
M/F}-theory compactification \cite{AW}. The intersection of D6
branes is purely geometrically described as unwinding of $U(m+n)$
singularity\footnote{We require a more Abelian part than a simple
$A_{m-1}$ singularity, to achieve the unwinding \cite{KV}, which
cope with the intersecting brane picture \cite{BVS}.} to those of
$U(m)$ and $U(n)$ and the bifundamental fermions come from the
branching of the adjoint
$$ \bf (m+n)^2 = (m^2,1)+(1,n^2)+(m,n) $$
In the intersecting
brane picture, a string stretched from $m$ brane stack to $n$ brane
stack corresponds to {\em chiral} bifundamental representation $\bf
(m,n)$, and the other string having the opposite orientation is
$CPT$ conjugate due to the opposite GSO projection \cite{BDL}. We
will see that the class of vacua we will describe matches very well
with $M$-theory picture, since we have only six dimensional objects,
O6 and D6, which will lift to the above singularities.

\subsubsection*{A toy Madrid model}

A typical example is toy ``Madrid Model''. The gauge group
$H=U(3)_C\times U(2)_L \times U(1)_R \times U(1)_N$ arises from
breaking an unified group $G=U(7)$ whose adjoint have the following
charge assignment under the above subgroup $H$,
\begin{equation} \label{su7adj} {\bf 48} =
\begin{pmatrix}
\bf 8 & q & u & \\
q^{\rm c}   & \bf 3 &   & l \\
u^{\rm c}   &   &  \bf 1 & e \\
   & l^{\rm c}  & e^{\rm c}  & \bf 1 \\
\end{pmatrix}
\end{equation}
where the notations are self-explanatory; the block-diagonal
numbers refer to the dimensions of gauge bosons, and off-diagonal
blocks correspond to complex squarks and sleptons and hermitian
conjugates. We will see the fields corresponding to blank entries
are massive. They have the correct quantum numbers, as well as
hypercharge defined as linear combinations
\begin{equation} \label{ycharge}
 Q_Y = \frac13 Q_C - Q_L - Q_R.
\end{equation}
This breaking is achieved by the following background magnetic
flux $F = 2 \partial_{[4}A_{5]}$ on the extra two-torus $T^2$,
\begin{equation} \label{toyBG} \begin{split}
 2 \pi \alpha' A_5 =
 \begin{pmatrix} {m_1 \over n_1} \1_{n_1} & & &  \\ &
 {m_2 \over n_2} \1_{n_2} & & \\ & &  {m_3 \over n_3} \1_{n_3} & \\ & &
 &  {m_4 \over n_4} \1_{n_4}
 \end{pmatrix} x^4
 +
 \begin{pmatrix} a_1 \1_{n_1} & & &  \\ &  a_2 \1_{n_2} & & \\ & &
 a_3 \1_{n_3} & \\ & &  &  a_4 \1_{n_4}
 \end{pmatrix}
.
\end{split}
\end{equation}
Here gcd$(n_a,m_a)$ are $3,2,1,1$, respectively. In compactification
with more dimensions, we may put some of the blocks to different
gauge field components than $A_5$.

The example (\ref{su7adj}) thus is obtained as follows, where
$T$-dual picture is more transparent. Start with 7 slices of
coincident branes, separate stacks with $3_C+2_L+1_R+1_N$ branes.
Then after rotating $2_L$ and $1_N$ branes, we obtain chiral
spectrum. One can check \cite{GR} that the commuting generators of
$U(7)$ represents unbroken gauge group and chiral quarks and
leptons, exclusively not paired with the charge conjugate,
indicated in (\ref{su7adj}).

In the effective field theory level, two ingredients made the above
physics possible: extra dimensions and constant magnetic flux. This
would motivate model construction as an effective theory.

\subsection{Orientifold plane and mirror cycle}

The above toy model is anomalous, which is not bad since we have not
considered the complete theory yet. The best explanation for an
anomaly free theory is that it is a spontaneously broken phase of a
unified theory, in which the absence of anomaly is natural. The most
suggestive scenario will be minimal, ten dimensional supergravity
with nonabelian gauge group with dimension 496 \cite{GS}. This
suggests type I string theory with the gauge group $SO(32)$, which
is the only possibility if we want the perturbative open string
description.

In the brane picture, consistency is given by RR tadpole cancelation
condition. It guarantees anomaly freedom of nonabelian gauge
anomalies and of other mixed anomalies involving $U(1)$'s by the
generalized Green--Schwarz (GS) mechanism from antisymmetric tensor
fields \cite{GS,BM}. For this we introduce an orientifold plane
(O-plane), a fixed plane under both worldsheet parity reversal
$\Omega$ and spatial reflection. The closed string exchange between
D-branes (represented by cylinder Feynman diagram) give rise to
divergence, which is represented as tadpole diagram in the filed
theory limit \cite{PC}. The cylinder diagram is canceled by
introducing the Klein bottle and the M\"obius strip, which means
nothing but the existence of an orientifold plane with RR charge 16
in terms of D-brane unit. An antibrane can be introduced to cancel
tadpoles, but it spoils all the supersymmetry and brings about
instability.

When branes (and their images) are on top of orientifold plane, the
gauge group is $SO$ type. The symmetry breaking $SO(A)\times U(B)
\subset SO(A+2B)$ is again described by parallel separation away
from an O-plane. The adjoint $A$ of $SU(N)$ can be embedded into
$SO(2N)$
\begin{equation} \label{so32embedding}
 \begin{pmatrix} A & \\ & -A^\top \end{pmatrix}
\end{equation}
For example, the above $SU(7)$ toy model is realized and embedded in
$SO(14) \subset SO(32)$, having almost same structure. Since the
final group is $SO(32)$, so that in general we expect hidden sector
\cite{Ni}. We will see also that, additional projections associated
with orbifold action breaks such $SO(N)$ group to $Sp(k)$.

The ``mirror'' matrix $A^\top$ in (\ref{so32embedding}) represent
the mirror image with respect to orientifold plane, as follows. An
orientifold action having $(1,0)$ orientifold plane maps $(n,m)$
cycle to $(n,-m)$, and one with $(0,1)$ maps $(n,m)$ to $(-n,m)$.
This brings about mirror fermion having charge $\bf (m,n)$ for $\bf
(m,\overline n)$ (or $\bf (\overline m, n)$ depending on the
orientation) corresponding to mirror cycles. For a six dimensional
cycle (\ref{18cycle}) we have the mirror cycle
\begin{equation} \label{18mirror} \begin{split}
 & (n_1,-m_1)(n_2,-m_2)(n_3,-m_3) \\ & = (n_1 n_2 n_3, n_1 m_2 m_3, m_1 n_2
  m_3, m_1 m_2 n_3, -m_1 m_2 m_3, -m_1 n_2 n_3, -n_1 m_2 n_3, -n_1 n_2
  m_3) \\
 & = (N,c^{12}_2,c^{34}_2,c^{56}_2,-c_3,-c_1^{12},-c_1^{34},-c_1^{56}).
\end {split} \end{equation}
They are images with respect to any of the
following O-planes
\begin{equation} \label{z2z2oplane}
 (1,0)(1,0)(1,0), \quad (0,1)(0,1)(1,0), \quad (1,0)(0,1)(0,1), \quad (0,1)(1,0)(1,0)
\end{equation}
which are the fixed planes of orientifold action $\Omega R, \Omega R
\theta_1, \Omega R \theta_2, \Omega R \theta_1
  \theta_2$, in (\ref{z2z2act}) and (\ref{Ract}).
Tere are only two orientation for a D-brane, thus flipping some signs
does not affect actual cycle.

For tilted torus discussed above, the situation is the same. We have
two choices for each two-torus. A $\Z_2$ rotation relates them but
it does not simply rotate. The rank is reduced. Correspondingly, the
image of D-brane with respect to such plane is
\begin{equation}
 (n,\tilde m) \leftrightarrow (n,-\tilde m).
\end{equation}

This will set an upper bound of the rank of the gauge group.
 This property actually exists since the
tadpole cancelation condition is the sum of RR charge vanish
\begin{equation} \label{tadpolecond}
  \sum_a N^{a} = \sum c_2^{ij,a} = 0 \quad \text{(including O-planes)}
\end{equation}
including orientifold plane. These are four conditions since we have
3 complex dimensions thus $ij$ runs over $45,67,89$. We will see
that there are three cases that one, two and four of them is
nonvanishing which is the same as the number of orientifold planes.

\section{Connected vacua}

As discussed in introduction, an obvious flat direction corresponds
to brane separation, and it breaks or enhances the gauge symmetry
without energy cost. We argue here another flat direction which
affects intersections.

\subsection{Supersymmetric cycles}

If we have two branes at the same time, each preserves
different components of the supersymmetry and the surviving
supersymmetry is their
intersection \cite{BDL}
\begin{equation}
  Q + \beta \tilde Q \text{ and }  Q + \beta' \tilde Q = Q + \beta' \beta^{-1} \beta \tilde Q,
\end{equation}
of two supersymmetries generated by $Q$ and $\tilde Q$ and $\beta$
and $\beta'$ are a projection in the transverse directions. $\beta'
\beta^{-1} = e^{i \sum \theta_i J_i}$, whith $J$ a rotation
generator on spinors, parameterized by angles $\theta_i$ along an
$x$-axis for each $i$th two-torus $T^2$. The surviving supersymmetry
is the common intersection such that
\begin{equation}
 \beta' \beta^{-1} =  1.
\end{equation}
This is equivalent to
\begin{equation} \label{18susy}
\pm \theta_1 \pm \theta_2 \pm \theta_3 = 0 \Leftrightarrow  \pm
f_{45} \pm f_{67} \pm  f_{89} = (\pm f_{45})(\pm f_{67})(\pm
f_{89}),
\end{equation}
which comes from the addition formula for tangent functions and the
signs on each side are correlated. We denoted the backgrounds
$f_{mn} \equiv 2\pi \alpha' \langle F_{mn} \rangle$.
 The number of zero eigenvalue
$\theta_i$ determines how many supersymmetries remain. This also
shows that the same components of supersymmetry is preserved if
cycles are connected with the same holonomy group element $\beta'
\beta^{-1}$. This condition should be a relative relation between
two branes. The most convenient choice is to fix orientifold plane
and rotate D-branes by holonomies, which guarantees the holonomy
relation between orientifold images.

This condition plays the role of F-flat condition \cite{MMMS}. From
the index theorem (\ref{diracidx}), the fermions are always
massless, so that this supersymmetry condition is equivalent to how
many scalar fields can be their massless superpartners. Also note
that the above condition requires some of the moduli fixed: Two
cycles connected by $SU(3)$ holonomy but not by $SU(2)$, fixes all
the complex structures.

\subsection{Born--Infeld energy}

The worldvolume theory of D$p$-brane is described by low energy
fields via DBI action
\begin{equation} \label{DBI}
 S = - T_p \int d^{p+1} x \Tr e^{-\Phi}
  \sqrt{-\det(G_{ab}+B_{ab}+2\pi \alpha' F_{ab})},
\end{equation}
with the tension $T_p = 2 \pi (4 \pi^2 \alpha')^{-(p+1)/2}$.

\begin{figure}[t]
\begin{center}
\includegraphics[height=2.5cm]{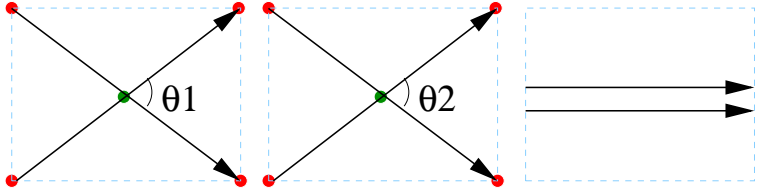}
\end{center}
\caption{1/4 BPS cycle. Only the 1/4 cycle $\theta_1 = \theta_2$ can
make supersymmetric state and recombine without energy cost. The two
cycles are mirror image with respect to
  the orientifold plane.} \label{fig:14int}
\end{figure}

\begin{figure}[t]
\begin{center}
\includegraphics[height=2.4cm]{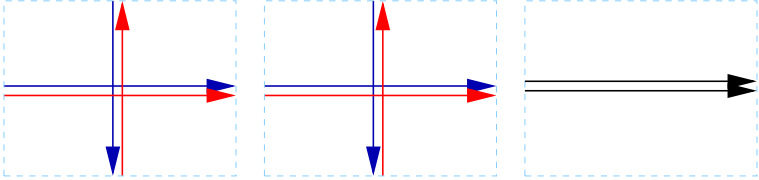}
\end{center}
\caption{A the branes are on top of orientifold planes: A
phase obtained from type I compactification. There is extra gauge
  symmetry, not visible in gauge thoery, supperted by instantons.
  }
\label{fig:14parallel}
\end{figure}

First focus a constant flux configuration, provided by a D-brane
background, satisfying the supersymmetry condition (\ref{18susy}) and
always dualized to a D9-D5 system. All
the flux lie in abelian subgroup, thus commute. Among four
possible combinations, we take all the sign to be positive for convenience.
Plugging it to (\ref{DBI}) we have the energy relation \cite{MMMS}
\begin{equation} \label{energies} \begin{split}
H =&\ \tau_9 V_2 V_3 V_4 \sum_{a=1}^k \Tr \big[(\1_{N^a}+
 f_{45,a}^2)(\1_{N^a}+ f_{67,a}^2)(\1_{N^a}+ f_{89,a}^2) \big]^{1/2} \\
 =&\ \tau_9 V_2 V_3 V_4 \sum_a \Tr \bigg[ (\1_{N^a}-f_{45,a} f_{67,a} -
 f_{67,a} f_{89,a} - f_{89,a} f_{45,a})^2 \\
 & \qquad +
(f_{45,a} f_{67,a} f_{89,a} -f_{45,a} - f_{67,a} -f_{89,a})^2 \bigg]^{1/2} \\
 =&\  \tau_ 9 V_2 V_3 V_4 \sum_a \left( \Tr \1_{N^a}- \Tr f_{45,a} f_{67,a} - \Tr f_{67,a} f_{89,a} -
 \Tr f_{89,a} f_{45,a} \right) \\
 =&\ \tau_9 V_2 V_3 V_4 \sum_a N^a - \tau_5 \sum_a ( V_4 c_{2,a}^{89} +
 V_2 c_{2,a}^{45} + V_3 c_{2,a}^{67}).
\end{split} \end{equation}
Here $k$ is the number of stacks. We stress the negative signs in
front of $c_2$ are dependent on the sign convention in (\ref{18susy}),
which will be discussed shortly.
We also used the dilation relation (\ref{tension})
and the toron quantization condition (\ref{4quant})
\begin{align*}
  \tau_5 &= (4\pi^2\alpha')^2 \tau_9 \\
  V_2 V_3 \Tr F_{45,a} F_{67,a} &= V_2 V_3 \Tr
(F_{45,a})^2 = N^{a} {m_1^{a} m_2^{a} \over n_1^{a}
 n_2^{a}} = c_2^{a}.
\end{align*}
The DBI energy is expressed
in terms of the linear sum of the charges. This is the realization of typical
property of BPS saturated states and the relation is exact in
$\alpha'$ expansion.

Now, this energy is constrained by the RR tadpole cancelation
(\ref{tadpolecond}).
\begin{equation}
 \sum N^a =  16, \quad \sum c_{2,a}^{ij} = \text{ fixed}
\end{equation}
including orientifold mirror images. Definitely the matrix $f_{mn}$
should be embedded $SO(32)$ gauge group. The sum of the rest charges
$\sum_a c^{ij}_{2,a}$ is dependent on the number of extra O5 planes,
which is determined by orbifold. The maximal number of orientifold
planes, or disconnected gauge group is therefore four.

We thus have arrived at the important observation:
Every consistent system has {\em the same energy because of the tadpole
condition}. Furthermore, we have supersymmetry, which has many flat
directions, thus we expect that many {\em vacua are connected.}
Among the possibilities, our interest is that from {\em type I
compactification}, where all the D-branes are on top of orientifold planes.

\subsection{Brane recombination} \label{sec:recomb}

We will seek the possibility that a given intersecting brane setup
is continuously connected to a type I compactification. This is done
by brane recombination: branes can be deformed yielding to another
configuration.

Consider 1/4 BPS cycles first, corresponding to one vanishing angle
$\theta_3=0$ case as in (\ref{t2cherns}). $T$-dual in 579 directions
makes it to D9-branes with magnetic fluxes, which is equivalent to
bound D5-branes filling in 89 directions. By a rotation, we can
always make the magnetic fluxes into block diagonal and diagonal in
the spacetime and the group space, respectively. Therefore, the DBI
energy (\ref{energies}) is
\begin{equation} \label{14dbi}
   H =  \tau_9 V_2 V_3 V_4 \sum \Tr (\1_{N^a} + f^2_{45,a})
   =  \tau_9 V_2 V_3 V_4 \sum N^a + \tau_5 V_2 \sum c_{2,a}
\end{equation}
The sign of the second term is always positive for any choice of the
sign.

From worldsheet bosonic spectrum, we have two light complex scalars
with masses
\begin{equation} \label{tachyons}
  M_1^2 = {1 \over 2 \pi \alpha'}( \theta_1^{a} - \theta_2^{a})
 = - M_2^2
\end{equation}
where $\theta_i^{a}$ is the angle of 2-cycles $\pi_a$ in $i$th
$T^2$. We have a tachyon for generic angles $\theta_1^{a} \ne
\theta_2^{a}$, which reflects the instability of vacuum, arising
from wrong expansion of the (string) field theory. This tachyon
condenses and rolls down to a true runaway vacuum \cite{Sen}.
However, if
\begin{equation} \label{su2hol}
\theta_1^{a} = \theta_2^{a} \quad \text{ for all }a,
\end{equation}
there is no tachyon. This demonstrates  that we
have supersymmetry, since the superpartner fermions are always
massless. This condition is translated into the condition
on magnetic field
\begin{equation}
 F_{45}^{a} = \pm F_{67}^{a} \Leftrightarrow  *_4 F_{mn}^{a} = \pm
 F_{mn}^{a}.
\end{equation}
Although we have made use of factorizable cycle, the latter
(anti-)self-dual condition is general for nonfactorizable cycle
\cite{BT} $m,n=4,5,6,7$. The Hodge-dual operator is in the
4567-direction. Then the preserved supersymmetry
component\footnote{The tilted 1/2 cycle, which is not self-dual, can
be supersymmetric, if (\ref{selfdual}) is supplemented by an
additional term from $U(1)$ factor \cite{GGBL,GR97}. However in the
presence of more than one supersymmetric cycles, the nontrivial
setup cannot make use of such additional term.} $\epsilon$
\begin{equation} \label{selfdual}
 \delta \chi = \Gamma^{\mu \nu} F_{\mu \nu} \epsilon = (\Gamma^{45} \pm
\Gamma^{67} ) F_{45} \epsilon =  0
\end{equation}
is same for every intermediate step. Since supersymmetry is
preserved, there is no cost on ``marginal'' deformation. Therefore
we see, for 1/4-cycle, there is a recombination \cite{Choi06}
\begin{equation*} \begin{split}
&(n_1,m_1)(n_2,m_2)(1,0)+(n_1,-m_1)(n_2,-n_2)(1,0) \\
 &\to  2n_1 n_2(1,0)(1,0)(1,0)+2m_1 m_2(0,1)(0,1)(1,0).
\end{split} \end{equation*}
which preserves the energy and charge. The two phases
are drawn in Figs. \ref{fig:14int} and \ref{fig:14parallel}. (The
parallel phase cannot be described by gauge theory, because of the
vertically running brane.)

\begin{figure}[t]
\begin{center}
\includegraphics[height=3.2cm]{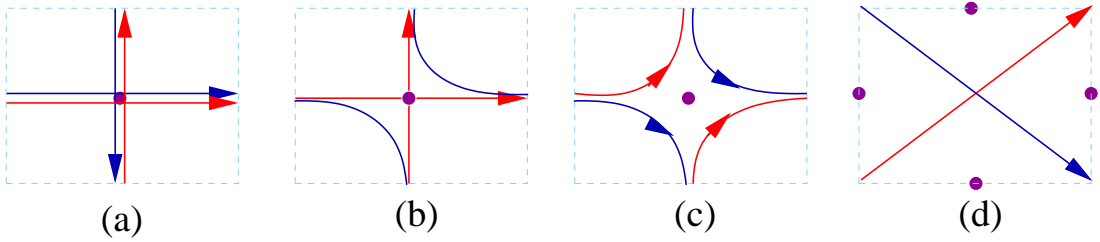}
\end{center}
\caption{Intermediate steps, governed by the supersymmetry condition
(\ref{18susy}), which is a generalization of self-dual condition.
The points denote intersections, some of which are identified.
During the recombination the RR charges and the (sum of) volumes are
conserved. Only one two-torus is drawn.} \label{fig:restor}
\end{figure}

The recombination of 1/4 cycles is well known as fattening and
shrinking instanton. We take a simplified example of bound state of
two D5 and two D9 branes in a rectangular $T^4$ in 6789 direction.
When we make a $T$-dual in 68-direction, the D5 and D9 branes become
two D6 brane at right angles, as in Fig. \ref{fig:14parallel}.
Definitely the degree of freedom is $U(2)\times U(2)$. The gauge
theory cannot describe this, because in the original picture, there
is an extra, singular $U(2)$ gauge symmetry supported by instanton
\cite{Wi95}. A string between D5 and D9 brane transforms as
bifundamental $\bf (2,2)$ denoted $h_M^{Am}$ in ref. \cite{WiDo}:
$m$ and $M$ are internal index of D5 and D9 gauge theory,
respectively, and $A$ is R-symmetry index $SU(2)_R$ as a part of
$SO(4)$ internal Lorentz symmetry transverse to D0-brane. $h^A$
corresponds to a Higgs field, with the charge parameterized by $m$
and $M$, and describes the scale size of the instanton. Among them,
there is a setup in which we have two tilted intersecting brane,
depicted in Fig. \ref{fig:14int}. We see the resulting two brane has
separated by the diagonal distance of the torus, as in Fig.
\ref{fig:restor} (b) or (c). The intersection point is specified by
D0-D0 state denoted as $X^{AY}$ in \cite{WiDo}, where $Y$ is
leftover $SU(2)$ index of above transverse $SO(4)$. We can
generalize this to any of D5-D9 bound state.

Although the above phenomena is not realized by gauge theory, a
further Higgsing can be \cite{CIM02}. A bifundamental
representation, corresponding to off-diagonal elements in
(\ref{bifurep}), further {\em breaks} the ``visible'' gauge symmetry
and reduces the rank.
The electroweak Higgs mechanism is a good example. Note that,
for the intersecting setup, turning on the constant VEV does not
affect F-term condition (\ref{18susy}) given by derivatives.
However to fulfill D-term condition, the $i$th block of the coordinate
matrix $X^m_a$ should be either diagonal or
proportional to others. The former condition is responsible for
parallel translation discussed in introduction. The latter
completely determines the form of $X^m_a$ if the one component is
determined. In this sense the supersymmetric deformation is encoded
in one dimensional degrees of freedom $b^1(\pi)$. For example, we
have parameterized the 1/4 cycles (\ref{t2cherns}) as (\ref{4quant})
and it is the slope of tilted branes, as in (\ref{growvev}),
\begin{equation}
 \frac 1{2\pi\alpha'} X^2_a = {2 \pi \over A}  {m^a_1 \over n^a_1}
  \1_{N^a }X^1  + A^0_{2,a}
\end{equation}
Due to off-diagonal elements, we lost the geometrical
interpretation. Being Hermitian, we seems to be able to recover the
geometry if we diagonalize all the $X^m$ simultaneously, this would
not be compatible with the torus embedding. Anyway, in the
noncompact limit, or the local limit around the intersections,
diagonalization gives curves in general, also look like Fig.
\ref{fig:restor}. Also in $T$-dual space, the magnetic flux is not
uniform which is possible because both the instanton field and the
energy scale the same. In the limit of (at least one) off-diagonal
components of $X^2$ goes to large, the finite $X^2$ dependence goes
to zero, leaving the parallel branes. This is again the same kind of
Higgs mechanism, and the states corresponding to off-diagonal
element transform as ifundamental representation. Of course, every
intermediate curves has the same volume since the constant Wilson
line is only responsible for translation; In the $T$-dual picture
the DBI action gives the volume itself, as a function of
derivatives. We can relate this Higgs mechanism with D-term
supersymmetry breaking.

Describing tilted or intersecting brane does not require any further
moduli \cite{BDL}. It is determined by internal geometry, here torus
with two complex moduli, and quantization condition
(\ref{toronqtz}). Therefore the recombination is the only possible
way to change intersecting setup involving chirality: There can be a
{\em local change in chirality} \cite{DZ}, as clearly seen in Fig.
\ref{fig:restor}, and the intersection number (\ref{intersecnum}),
but the {\em gross} chirality does not change since the RR charges
is invariant at every step. We can see this from the fact that Chern
numbers are not changed thus does not affect total anomaly
polynomial, reflecting again the fact that the anomaly is controlled
by RR charges.

We have some other descriptions.
The conditions (\ref{su2hol}) and (\ref{selfdual}) are
local relations \cite{BBH}. It is nothing but the holomorphic
Cauchy--Riemann condition
\begin{equation}
 {\partial x^5 \over \partial x^4} =  {\partial x^7 \over \partial
 x^4}, \quad
  {\partial x^5 \over \partial x^6}= -{\partial x^7 \over \partial x^6}.
\end{equation}
We parameterize the curve as $x^5+ix^7$ as a function of $x^4+ix^6$.
The first is equivalent to Eq. (\ref{su2hol}), follows from $\tan
(\theta_1-\theta_2) \propto \tan \theta_1 - \tan \theta_2=0$. Also
if we require the rotation to be unitary, we obtain the latter. The
complexified function $x^5 + ix^7$ is analytic and this fact
guarantees the holomorphicity in entire domain. It is well known
that complex curve in two complex dimension always span the minimal
volume \cite{Jo}, agreeing with (\ref{energies}). To the other way
around, since in two dimension, a holomorphic transformation is conformal
transformation, it always preserves relative angles (\ref{su2hol}) at
every point. Such property of minimality is known as calibration
between two cycles, and the corresponding (relative) cycle is
called special Lagrangian (SL) cycle \cite{Jo}. In ref.
\cite{BVS,DZ} an algebraic description of such interpolating curve
is presented. Similarly we have the anti-holomorphic condition from
anti-self duality.

We saw that supersymmetry-preserving
deformation is parameterized by Wilson lines of one coordinate
$X^m$, and this provides the special case of McLean's theorem,
applicable to the former case also \cite{Mc}: For a compact SL cycle
$\pi$ in an (almost) Calabi--Yau manifold, the moduli space of
deformation of $\pi$ is a smooth manifold of dimension $b^1(\pi)$,
where $b^1(\pi)$ is the first Betti number of $N$.

The above D0-D4 bound state is $U$-dual to (F,D$p$) bound state
which is interpreted as the electric flux on D-branes \cite{HT,CK}.
The above BPS equation becomes string ``junction condition'' \cite{DM}
(In the self-dual case $g=1$.) and we can clearly see the
supersymmetry condition at every local point, with the same
supersymmetry components preserved. In this picture we can also see
the no-energy-cost property under marginal deformation and the
specific shape of final state \cite{CK}. The DBI energy is $V\sqrt
{-\det {(1+2 \pi \alpha' F)_\mu}^\nu}.$ When we have purely magnetic
flux, this is nothing but the winding volume of torus
$V\sqrt{1+(2\pi \alpha' F)^2} = V\sqrt{1+\tan^2 \theta} = V_{\rm
cycle}$, agreeing with intersecting brane case. But when we have
contribution from fundamental string, which carries NS-NS flux,
which is equivalent to electric field component. The size of
electric and magnetic fluxes should be equal due to BPS condition,
so the determinant becomes 1, which shows independence of cycle.
Although there are not always one to one correspondence among dual
theories, at least it exhibit the property that there is no volume
dependence.

The case of 1/8-cycles is less trivial. From the dependence on
$N,c_2^{ij}$, we see that the system is $T$-dual to D9-\D5-\D5-\D5
bound state.\footnote{A D9-D3 bound state, and so on, is possible if
we turn on NSNS $B$ field \cite{Wi20}.} Three \D5-branes span along
89,45, and 67, directions. Protected by supersymmetry, the bound
state energy is again marginal. Not like the 1/4 BPS cycle case,
here the energy is dependent on the choice of the signs in
(\ref{18susy}), but the physics is the same. The system at hand
seems like $T$-dual to D3-D9 bound state, since one D6 becomes
pointlike in six compact dimension while the other becomes D9,
however it is not. A D3 charge would appear as a single $c_3$ and
D3-D9 system has only one sector having positive zero point energy
thus cannot be bound state. The D9-\D5-\D5-\D5 bound state has four
disconnected sector, only one of which has lowest mass zero and
other three lowest masses are positive. This is the consequence of
the supersymmetry condition (\ref{18susy}). Note also we can flip
some of the antibranes into D-branes, which amounts to flipping some
of the signs of the fluxes $f_{mn} \to - f_{mn}$. However we can
veryfy that not all of them can be. This property shows the reason
why this system cannot be reduced into D9-D5 systems. Being
quadratic, the leading order kinetic terms are invariant.

Still, there is a recombination, since for generic angles the system
has a tachyon, which again signals the recombination, and the
supersymmetric condition become the one of marginal stability. As in
the 1/4 case, VEVs of all the \D$5_i$-D9 Higgs is will describe
recombination, however, now the condition (\ref{18susy}) relates all
of them and thus is more strict. At the moment there is no concrete
solution to 1/8 case, from a general intersection to the trivial one
where every D-branes are on top of orientifold plane. However by
energetics and charge conservation, this is highly likely. Finding
the generalized instanton solution for 1/4 cycle \cite{MMMS} would
be interesting investigation.

\section{Gauge coupling unification}

We saw that each simple or Abelian subgroup came from the breaking
of a single unified group. Thus it might be surprising if each the
subgroup has a different gauge coupling. However we will see that,
because of the nonlocal property of DBI action, the background flux
can give rise to different factor for each subgroup, leading to a
different four dimensional gauge coupling.

\subsection{Gauge coupling}

Expanding to the quadratic order in $\alpha' F_{\mu \nu}$, the DBI
action (\ref{DBI}) reduces to ($p$+1)-dimensional Yang--Mills (YM)
action,
\begin{equation} \label{derivedYM}
 S_p = - {T_p (2\pi \alpha')^2 \over 4 g}
  \int d^{p+1} x ~\Tr F_{\mu \nu} F^{\mu \nu}
\end{equation}
where $g=e^{\langle \Phi \rangle}$ is the string coupling.
We are using the trace in the fundamental representation of $U(N)$.
Therefore, the YM coupling is
\begin{equation}
 g^2_{p+1,\rm YM} = (2\pi)^{p-2} g \alpha^{\prime (p-3)/2}.
\end{equation}

For the action to be invariant under $T$-duality, the dilaton should
transform under $T$-duality. Recover the dilation dependence and
consider a compactification in $p$-direction with the circumference
$L_p$. Compare the mass of D$p$-branes and its $T$-dual
D$(p-1)$-brane, which are the same,
\begin{equation}
  T_p e^{-\Phi} L_p = T_{p-1} e^{-\Phi'}.
\end{equation}
For them to be the same, the dilatons should behave as
\begin{equation}
  e^{\Phi'} = {\alpha^{\prime 1/2} \over L_p} e^{\Phi}.
\end{equation}
Hence the gravitational coupling and gauge coupling, which are
proportional to $e^{-\Phi}$
It will be convenient later to define
\begin{equation} \label{tension}
 \tau_p \equiv T_p e^{-\Phi} = \tau_{p-1} L_p^{-1},
\end{equation}
since $\tau$ times the volume occupied by D-brane is an invariant of
$T$-duality. This is nicely expressed as the relation between
gravitational and gauge coupling
\begin{equation} \label{type1couprel}
 {g_{\rm YM}^2 \over \kappa}= 2 (2\pi)^{7/2} \alpha'
\end{equation}
where the factor 2 with a single orientifold plane, for any number
of compact extra dimensions. When all the branes are on top of
orientifold plane, this is the same relation to that of type I
string.

The tilted D-brane is represented by a constant magnetic flux. Now
we calculate the actual fluctuation around D-brane background
\cite{AHT,DST}
\begin{equation} \label{fluc}
A_\mu = A_\mu^0 + \delta A_\mu.
\end{equation}
In fact the expansion is more involved, since fluctuations do not
commute with background $ [ A_\mu^0 ,\delta A_\mu] \ne 0. $
For nonabelian DBI action, the trace in (\ref{DBI}) should be done
on the group space. Because the noncommutivity, how to evaluate the
trace is not known so far.  Replacing trace with symmetrized trace
\cite{Ts} $ \Tr \to \STr $ works very well. This is known not fully
consistent remedy, since at the quadratic order, spectrum of
intersecting brane does not match with the string calculation
\cite{BBRS,AHT}.

Plugging (\ref{fluc}) into the DBI action (\ref{DBI}) and expanding
up to the quadratic order in $\tilde F_{\mu \nu} \equiv 2
\partial_{[\mu} \delta A_{\nu]}-i[A_\mu,\delta A_\nu]$, around
\begin{equation}
 {(\1 + f){_\mu}^\nu}.
\end{equation}
The raising and lowering indices are done
with the `genuine' flat metric $\eta_{\mu \nu}$. Here we keep in
mind that this has both the Lorentz and the gauge group index, and
we suppressed the latter. We can decompose its inverse to a
symmetric and antisymmetric part in Lorentz index,
\begin{equation}
  (\1 + f)^{-1} = (\1-f^2)^{-1} - f (\1-f^2)^{-1} \equiv g + b,
\end{equation}
where the inverse is done on both the Lorentz and the gauge group
space. The DBI action becomes
\begin{equation} \label{DBIsep}
  \STr \sqrt{\det(\1+f)} \sqrt{-\det(\1 + g \tilde F + b \tilde F)}
\end{equation}

\subsection{Fluctuation spectrum}

We expand the action (\ref{DBIsep}) to the quadratic order in
fluctuation $O(\alpha' \tilde F)^2$. The leading order constant term
is the background contribution (\ref{14dbi})
\begin{equation}
S = - \tau_9 \int d^{10} x \Tr \1_{32} -{1 \over 4 g_{\rm YM}^2 }
\int d^{10} x \Tr F_{MN}^2 + \cdots.
\end{equation}
The gauge group for the first term is always $SO(32)$ with type I
coupling $g_{\rm YM}$ (\ref{type1couprel}), since this is the
expectation value of a constant magnetic flux in type I string
theory. If we have more orientifold plane(s), there can be other
nonzero second Chern numbers, thus be as many $SO(32)$, denoted by
ellipsis. However all the couplings are the same, since the
orientifold planes and the D-branes on top of them are exchangeable
by a $T$-duality (which is not visible in gauge theory).

Now consider the fluctuation around the above background---of the
1/4 cycle first. We have
\begin{equation} \label{DBIexp}
  L = \STr (\1 + f_{45}^2)
 \left( g_{\mu \nu} \tilde F^{\nu
 \lambda} g_{\lambda \sigma} \tilde F^{\sigma \mu} +
 b_{\mu \nu} \tilde F^{\nu
 \lambda} b_{\lambda \sigma} \tilde F^{\sigma \mu}
 \right) + \text{(topological terms)}.
\end{equation}
The total derivative term becomes topological term and contributes
to the zero point energy, agreeing with the worldsheet calculation
\cite{DST}. The nonvanishing contribution of $(\1+f)^{-1}$ is only
the symmetric part $g$ \cite{AHT} having eigenvalues
\begin{equation}
 g_{mn} =  \begin{pmatrix}   (1+f_{45}^2)^{-1} \1_2  & &
      \\  & (1+f_{67}^2)^{-1} \1_2 & \\ & & \1_2  \end{pmatrix}, \quad g_{\mu
      \nu} = \eta_{\mu \nu}.
\end{equation}
Thus we have the action
\begin{equation} \begin{split}
   2 \tau_9& (2\pi
   \alpha')^2  \int d^{10} x \STr (\1 + f_{45}^2)  g^{MN} \tilde F_{NP
} g^{PS} \tilde F_{SM}
\\
    &= S_{\rm 6D} - \frac1{4 g_{\rm YM}^2} \int d^{10}x \STr (\1 + f_{45}^2)  \tilde
   F_{ \mu \nu} \tilde F_{ \mu \nu} + \dots
 \end{split} \end{equation}
The D9 and D5 fluctuation is forced to be identified, which is
exactly the prefactor affecting the canonical normalization. For
012389 direction, the prefactor persists, however for 4567, which is
the brane-tilted space, the prefactor exactly cancel out the metric.

In the low energy limit where extra fluctuation is negligible
$\tilde F_{mn} \simeq 0$, we have four dimensional effective YM
theory
$$
   S= - \frac{v_6}{4 g_{\rm YM}^2} \int d^4x \Tr (\1 + f_{45}^2)  \tilde
   F_{ \mu \nu} \tilde F_{ \mu \nu} + \dots.
$$
with
\begin{equation}
 \1 + f_{45}^2 = \begin{pmatrix}
  (1+ (\frac {m^1} {n^1})^2)\1_{N^{1}} & & & \\
& (1+ (\frac {m^2} {n^2})^2)\1_{N^{2}} & &  \\ & & \ddots & \\
 & & & (1+ (\frac {m^k} {n^k})^2)\1_{N^{k}} \end{pmatrix}.
  \end{equation}
Note that, the $a$th gauge field come from the $a$th block of rank
$N^a$. The effective gauge coupling constant is
\begin{equation}
  g_{4,i}^2 = g_{\rm YM}^2/V_{{\rm cycle},i}
\end{equation}
from the canonical normalization. This is so because the gauge field
background described by DBI action multiplies the kinetic terms.
Since it has nontrivial nonabelian structure, it gives different
couplings to each subgroup. Being described by DBI action, a single
unified group $G$ can give different gauge couplings for each
unbroken factor groups $H_i$. Without introducing extra volume
moduli, nor introducing nonabelian structure to it, we can have
different gauge couplings for different gauge groups originating
from the same $SO(32)$.

Near and above the string scale $M_s \sim \alpha^{\prime -1/2}$, the
size of fluctuation becomes comparable to that of the background
$\delta A \sim A_0$, therefore the above expansion breaks down.
Above this scale also the extra dimensions open up, thus the theory
is ten dimensional type I string theory or at least its supergravity
approximation. This is the reason why the YM dynamics cannot catch
such dependence. However we can understand, at least, how
spontaneous symmetry breaking could lead a different coupling for
each gauge group at the breaking scale. This is explained by brane
recombination. Vacua with different four dimensional couplings are
connected by energy costless deformation, which is possible since
four dimensional coupling is controlled by dynamics of extra
dimension.

In the internal dimension, the canonical normalization is not
affected, thus the gauge coupling is the same. However in the
potential term, there is a rescaling by one factor of $g$
\cite{AHT}. We have used
$$ {(1 +  f)_\mu}^{\rho} {g_\rho}^\nu = \1 + \text{ (antisymmetric tensor)},$$
for the 45 and 67 components, where the latter does not contribute
in the Lagrangian, and the YM coupling relation.
$$
 S_{\rm 6D} = - \frac1{4 g_{\rm YM}^2} \int d^{10}x \STr  \tilde F_{0 m} \tilde
   F_{0 m}  - \frac1{4 g_{\rm YM}^2} \int d^{10}x
\STr \tilde F_{ik} g^{kl} \tilde F_{li}  \\
$$

The factor 2 comes from D9 and D5 contributions, and the resulting two
YM actions correlated.
For self-dual case, this action reproduces completely the same
spectrum as worldsheet analysis with the symmetrized trace
prescription
\begin{equation} \label{intprescrip}
  - \frac1{f_{45}} \int_0^{-f_{45}} df' g_{45} = - \frac 1{f_{45}}
    \int df' (1 + f_{45}^{\prime 2})^{-1} = {\theta_1 \over \tan \theta_1}
\end{equation}
and the metric dependence modifying momentum-winding quantum numbers.
The only difference is that we do not have overall factor in (\ref{DBIexp}).
This prescription correctly reproduces the spectrum, as obtained from
the worldsheet theory \cite{DST,AHT}.

%\subsection{Less supersymmetries}

Let us come to the 1/8 BPS cycles with no zero eigenvalues for
$\theta_i$. In this case, there is no known prescription to produce
the correct spectrum. We examine the reason here.
The leading zeroth order contribution (\ref{energies}),
which looks like instanton density, must be still the usual gauge
kinetic term with background value. Again, supersymmetry condition
(\ref{18susy}) plays a similar role as self-dual condition
\begin{equation} \begin{split}
  & f_{45} f_{67} + f_{67} f_{89} + f_{89} f_{45} \\
& = \frac12( - f_{45}^2 - f_{67}^2 - f_{89}^2 + (f_{45} + f_{67} +
 f_{89})^2) \\
& = - \frac14 f_{mn}^2 + \frac12 (f_{45} f_{67} f_{89})^2.
\end{split} \end{equation}
Note that we have recovered the correct sign for the kinetic term.
Therefore we have Hamiltonian density
\begin{equation}
  \frac1{4 g^2_{\rm YM}}  \int d^{10}x(   \Tr \1 +  \Tr F_{MN}^2 +
  { O}(\alpha' F)^4).
\end{equation}
Its fluctuation is expected to be give the desired gauge theory.
However there have no expansion of DBI action to the desired YM
action. We claim that the reason can be again tracked by nonlocality
of stringy physics at scale $\alpha'^{-1/2}$.

\begin{figure}[t]
\begin{center}
\includegraphics[height=4cm]{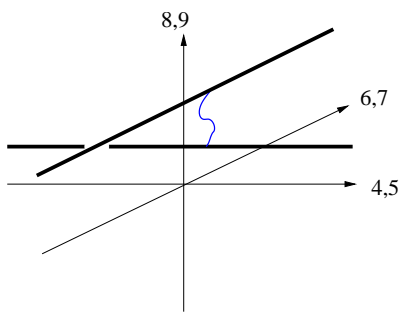}
\end{center}
\caption{Interbrane interaction is nonlocal, thus we look at the
  theory in the vicinity of a specific \D5 brane. Two \D5-branes are shown.}
  \label{fig:D5s}
\end{figure}

In the vicinity of 45 brane, we can see the effect of \D5-brane
along 45 direction, as gauge theory again. However we cannot see the
rest of gauge theory {\em here}, because of nonlocality: They are
cast as the higher order terms, because of the string stretched
between other \D5-branes along 67 and 89 directions. They are
suppressed by $(2\pi\alpha')^2$, which, by dimensional analysis,
shows that the interaction requires four more dimensions. Fig.
\ref{fig:D5s} shows that there are no intersection in this \D5-D9
picture and the interaction is nonlocal, which is the higher terms
of DBI action. In $T$-dual picture, in this 1/8 BPS case, the
magnetic flux due to D5 brane is not completely spread over the
transverse four dimension inside D9, therefore there is no
interaction among fluxes of different \D5 branes. However there is a
local relation with space-filling D9-branes with this \D5-brane,
being the rank $N$ gauge group. The permutation symmetry of the
terms, which is again the $T$-dual relation, shows that we have
similar effects around other \D5-branes.

What would be the non-supersymmetric coupling relation? We realize
that, the above supersymmetry relation is nothing but the BPS
inequality. For non-supersymmetric system, still the above action
(\ref{energies}) is the leading order terms, and the
non-supersymmetric term will be added to the higher orders from
${ O}(\alpha' F_{\mu \nu})^4$. Admittedly the BPS condition
(\ref{18susy}) make the supersymmetric expansion (\ref{energies})
exact. Also note that this relation holds for non-factorizable torus
and orientifold. Although in the nonsupersymmetric case it is not
clear that we have a similar condition to self-duality, we know that
D-brane plays the double role as instanton and gauge theory even in
nonsupersymmetric case. So it is reasonable to convert the
Chern--Simon $F^2$ terms to gauge kinetic ones and expect higher
order terms as non-local interactions.

We have briefly commented on a winding state by $w$ times, or bound
state of $w$ branes. From the former picture, it is evident that the
winding volume is $w$ times larger than the single one. However DBI
action cannot reproduce it, and in fact we have no tool taking into
account it.

\subsection{Weak mixing angle}

The conventional Grand Unification based on groups such as $SU(5)$ and
$SO(10)$ naturally give rise to nontrivial weak mixing angle
\begin{equation}
  \sin^2 \theta_W \equiv  {g_L^2 \over g_{\rm em}^2}
 = \frac 38
\end{equation}
at the unification scale. Within the current experimental limit,
this is the most desirable value. It is purely given by group theory
provided that there is a single gauge coupling. This is also
naturally realized in the heterotic string compactification
\cite{CK38}, although $U(1)$ factor is determined by model basis.

In the intersecting brane scenario, the natural gauge group is not a
simple but a semisimple group containing $U(1)$. For example
the $U(1)$ factor inside $U(N)$ is identified as the trace, and the
canonical normalization gives
\begin{equation} \label{U1canong}
 A_{\mu, U(1)} = \frac1N \Tr A_{\mu, U(N)}, \quad
 g^2_{U(1)} =  \frac2N g^2_{U(N)},
\end{equation}
for normalization $\Tr T^a T^b = \frac12 \delta^{ab}$ for
fundamental representation of $U(N)$. The hypercharge is given by
the linear combination of several $U(1)$s, $Q_Y = \sum_i C_i Q_i$.
According to term, canonical normalization of gauge kinetic term
yields,
\begin{equation}
  \frac1{g^2_Y} = \sum_i {N_i C_i^2 \over 2}  \frac1{g^2_i}.
\end{equation}
For the charge assignment (\ref{ycharge}), we have
\begin{equation} \label{Ymadrid}
  \frac1{g^2_Y} = \frac 16 \frac1{g^2_C} + \frac 1{g^2_L} +
 \frac12 \frac1{g^2_R} = \frac 53 \frac1{g^2},
\end{equation}
in the unified coupling limit $g = g_L = g_R = g_C$. We have the
desired relation
\begin{equation} \label{sinthetaw}
 \sin^2 \theta_W = {1 \over g_Y^2/g^2 + 1} = \frac 3 8
\end{equation}
 at the unification scale.

The reason for the correct value can be tracked
 to charge embedability into $SO(10)$. We note that in the Madrid model
 discussed above, the three anomalous extra $U(1)$ factors can be
 interpreted as baryon $Q_B$ and lepton $Q_L$ and Peccei--Quinn
 numbers $Q_R$
\begin{equation}
 A_{\mu,B} = \frac13 \Tr A_{\mu,U(3)_C}, \quad A_{\mu, L} = \frac12
 \Tr A_{\mu,U(2)_L}.
\end{equation}
These are broken by generalized GS mechanism, and become
global symmetries below the breaking scale.
This provides a good explanation of the accidental symmetries
of the baryon and the lepton numbers in the Standard Model, which remains
continuous. Higgsing with charged field always leaves the discrete
subset of the symmetry. The relation (\ref{Ymadrid}) comes from the
hypercharge relation in $U(1)_{B-L} \times U(1)_R \subset SO(10)$
\begin{equation}
  Q_Y = Q_B - Q_L -Q_R = \frac13 Q_C - Q_L -Q_R,
\end{equation}
reproducing (\ref{sinthetaw}). Obviously, matter spectra do not fit to
the spinorial ${\bf 16}$. However $SO(32)$ is
the group that can embrace all of these features. Note that, if we
obtained $SU(2)_L=Sp(1)$, rather than one above $SU(2)_L \subset
U(2)_L$, we could not reproduce the desired coefficient $1$ in the
second term on the (\ref{Ymadrid}), since $U(1)_L$ then would be an
independent Abelian gauge group having normalization $1/2$. Since
the baryon and the lepton numbers are conserved, there is no fast
proton decay and the neutrino is only Dirac.

Therefore in the intersecting brane scenario, the running of gauge
couplings from electroweak scale would not be changed provided the
coupling is unified. For one option, if two of the gauge group is
related by some symmetry such as parallel separation or reflection due
to orientifold, their couplings are exactly the same
\cite{BLS,Choi06}. The other possibility is that the discrepancy
might be approximately compensated by loop threshold corrections
\cite{LS}. As bottom-up approach, the possible values scenarios for
the desired weak mixing angle is analyzed in \cite{ADKS}.

\section{Type I compactification}

The modern viewpoint of type I string is type IIB string with one
O9-plane and 16 D9-branes \cite{Sa}. Introducing a D-brane requires
an orientifold plane with the same dimensionality for the
consistency. By $T$-duality, we can always convert these D-branes
and orientifold to 9+1 dimensional ones. Also, we have seen that by
recombination, we could make all the D-branes coincident on the
orientifold planes. Therefore we naturally expect that all the
intersecting brane models should be to type I theory compactified
  on orbifold or its $T$-dual, which we will argue here.

%\subsection{Symmetry breaking by orbifold}

The symmetry possessed by an open string is broken by projections
associated with elements in orientifold group $P_1 \cup \Omega P_2$
\cite{GP,AFIV,BGK}. We only consider a case having a geometric
meaning $P_1 = P_2 = P$. The action of an element $g \in P$ and
$\Omega h \in \Omega P$ act on Chan--Paton factor as
\begin{equation} \label{CPprojector} \begin{split}
 g &: |\psi,ij\rangle \to (\gamma_{g,p})_{ii'}|g\psi,i'j'\rangle
 (\gamma_{g,q})^{-1}_{j'j} \\
 \Omega h & : |\psi,ij\rangle \to (\gamma_{\Omega h,p})_{ii'}|h\psi,j'i'\rangle
 (\gamma_{\Omega h,q})^{-1}_{j'j}
\end{split} \end{equation}
where the second subscripts $p,q$ denote the dimensionality of branes.
We defined $\gamma_{\Omega h,p} \equiv \gamma_{h,p}
\gamma_{\Omega,p}$ since $\Omega h = \Omega \cdot h$ and
$\gamma_{k,p} = \gamma_{1,p}^k$ thus we have $\gamma_{1,p}^N = \pm 1$.
The surviving fields are invariant ones under the
projection (\ref{CPprojector}).

The orbifold group $P$, compatible with the lattice translations
defining the torus, should be a discrete subgroup of $SU(3)$
holonomy, to preserve at least $N=1$ supersymmetry in four dimension
\cite{DHVW,Heterotic}. An orientifold projection breaks half of the
supersymmetry. In the closed string sector, such a choice guarantee
at least one of the spinorial component of RNS or NSR state
invariant under holonomy. In the open string sector, this orbifold
images are connected again by the holonomy, so there is a compatible
$N=1$ supersymmetry. Such crystals are classified and finite in
numbers \cite{EK93,Heterotic}.

%\subsubsection*{Symmetric orbifold}

Essentially, type II string theory with a O-plane is ($T$-dual to)
{\em type I} string theory. We saw that there is at least one such
orientifold plane required. This can always be converted to O9 plane
by $T$-duality. In general, orientifold group is generated by
orbifold actions $g \in P$ and worldsheet parity reversals
$$
 \Omega \prod R_m (-1)^{F_L},
$$
 where $R_m$ are spatial reflections of
compact $x^m$ directions and $F_L$ is the spacetime fermion number.
This O$p$ plane is lying (or, defined) on the fixed plane under
$\prod R_m (-1)^{F_L}$. By the $T$-dual $\prod R_m (-1)^{F_L}$, we
can always convert it to $\Omega$, while orbifold group $P$ is
untouched. The type IIB theory with orientifold element $\Omega$ is
type I theory, so that every theory containing orientifold is
regarded as compactification of type I string theory with $P$. It
follows that the {\em orbifold group $P$} completely determines
additional orientifolds and thus gauge group. (We will consider an
asymmetric orbifold $\hat P$ shortly.)

Whenever we have a D-brane it is always $T$-dual to D9-brane, so
there is divergence proportional to the ten dimensional volume
factor $v_4 v_6$ \cite{GP}. It is canceled by introducing O9 plane
\begin{equation}
 {v_4 v_6 \over 16} \left \{ 32^2 - 64 \Tr (\gamma_{\Omega,9}^{-1}\gamma_{\Omega,9}^\top) +
 (\Tr(\gamma_{1,9}))^2 \right \}
\end{equation}
and 32 D9-branes and imposing
\begin{equation}
 \gamma_{\Omega,9}^\top = + \gamma_{\Omega,9}.
\end{equation}

In the standard compactification on symmetric orbifold $P$, in
addition to this O9, there can be other orientifold planes when the
orbifold group is compatible with $\Z_2$. Consider the $P=\Z_N$
orbifold first. For $N$ odd, the remaining divergence is
proportional to $V_4$ which is the regularized noncompact
dimensional volume. It should be finite otherwise we have no physics
to regulate in that direction. In this case, with a single
orientifold plane, it is shown \cite{BKLO} that there is no
nontrivial supersymmetric setup, except one where all the D-branes
are on top of orientifold plane. The constraint comes from the
tadpole cancelation condition of both RR and NSNS charges. Before
constraining tadpole condition, we have potentials for complex
structure moduli and dilaton. From the cancelation condition of
both, there arises an inequality which is only saturated only when
the complex structure makes all the D-branes on top of O-plane.

For $N$ even, the remaining divergence is proportional
\begin{equation} \label{O5div}
 {v_6 \over 16 v_4} \left \{ 32^2 - 64 \Tr (\gamma_{\Omega R,5}^{-1}\gamma_{\Omega R,5}^\top) +
 (\Tr(\gamma_{1,5}))^2 \right \}
 \end{equation}
to $1/v_4$, which manifestly shows that it is $T_{5678}$-dual to the
above divergence term, since it relates $v_4 \leftrightarrow
1/v_4,\gamma_{\Omega,9} \leftrightarrow \gamma_{\Omega R,5}$. This
means, introducing 16 D5-branes in these direction cancels the
divergence, and the requiring replacement $\gamma_{\Omega R,5}^\top
= + \gamma_{\Omega R,5}$ shows again the $T$-duality. $N$ even has
order 2 section and this harbors exactly such the O5 and D5 planes.
It follows
\begin{equation}
 \gamma_{1,9}^N = -1, \quad \gamma_{1,5}^N = -1
\end{equation}

There is an extra piece from the cylinder amplitude proportional to
$v_6$ only, which seems not has such $T$-duality, but it vanishes
for an appropriate choice of $\gamma$ basis. However other
divergences arising from $\Z_4,\Z_8,\Z_8'$ and $\Z_{12}'$ orbifolds,
proportional to $v_6/v_4$ cannot be canceled. So these choices of
orbifolds are not consistent for the symmetric orbifold
compactification. We will see, however, these orbifolds can be
consistent when we consider asymmetric orbifolds.

For $P=\Z_M \times \Z_N$ orbifold,
extra orientifold planes can emerge depending on whether such $\Z_M$
or $\Z_N$ contains order two element.
We will see that we can have either one or four orientifold planes in
total, if we take into account supersymmetry. The latter case is both
$M,N$ even and contains
the following $\Z_2 \times \Z_2$ orbifold group as a subgroup. It
is generated by $\{\theta_1,\theta_2\}$  where
\begin{equation} \label{z2z2act} \begin{split}
 \theta_1: (z_1,z_2,z_3) \leftrightarrow (-z_1,-z_2,z_3), \quad
 \theta_2: (z_1,z_2,z_3) \leftrightarrow (z_1,-z_2,-z_3).
\end{split} \end{equation}
Thus we have a nontrivial order 2 element $\theta_1\theta_2:
(z_1,z_2,z_3) \leftrightarrow (-z_1,z_2,-z_3)$. Each of them has a
5+1 dimensional fixed planes thus can harbor O5 plane.
 Still in this case, now the $5$-lanes
are labeled as $5_i$, each divergence condition is exactly the same
as (\ref{O5div}) for each $i$th orientifold direction \cite{BL}.

From these conditions, we can cancel the diagram and this give rise
to conditions on the projection matrices $\gamma$, and applying
these to (\ref{CPprojector}) we have broken gauge group. They are
determined without ambiguity. Note that this will be given as
branching rules. For example the compactification on the orbifold
$T^6/(\Z_2 \times \Z_2)$ \cite{BL}. And they are explained. There
exists a discrete torsion which produces a different form of one
loop amplitude \cite{DP}, we will not consider it here.

It follows that the $Sp(k)$ groups can only appear as broken
$SO(32)$, which takes into account small instantons \cite{Wi95}.

%\subsubsection*{Asymmetric orbifold}

For consideration of $T$-duality one should be careful to mention
$P$. It was noted in \cite{BGKL} that $T$-duality flips sign
asymmetrically to the left and the right mover,
\begin{equation}
  X_L = X'_L, \quad X_R = -X'_R
\end{equation}
if we do $T$-dual odd times to a two-torus parameterized by
\begin{equation}
  (z_L,z_R) \leftrightarrow (\bar z_L',z_R'),
\end{equation}
which is as good as complex conjugation in $T$-dual space. Then
the orbifold action
\begin{equation}
  g: (z_L,z_R) \to (e^{i \theta} z_L, e^{i \theta} z_R)
\end{equation}
 becomes asymmetric in $T$-dual direction
\begin{equation}
 \hat g = T g T^{-1} : (z_L',z_R') \to (e^{-i \theta} z_L', e^{i
 \theta} z_R')
\end{equation}
The cases of $\Z_2$ and $\Z_2 \times \Z_2$ are exceptions, since
$e^{\pi i}$ is just a change of sign, so asymmetric orbifold is the
same as symmetric orbifold.

The orientifold group is $\Omega P$, which changes into $\Omega R
P$ by $T$-duality in $R$ direction, whereas the orbifold group
remains as $P$. Consider $T$-duality in $579$ direction, which is
conveniently realized as
\begin{equation} \label{Ract}
 R: (z_1, z_2, z_3) \leftrightarrow (\bar z_1, \bar z_2, \bar
 z_3).
\end{equation}
Now the orientifold group is $\Omega R P$, whereas the orbifold
group $P$ remains invariant. It is easily checked that all the O9
and O5 planes becomes O6 planes,
\begin{equation}
 \Omega, \Omega \theta_1, \Omega \theta_2, \Omega \theta_1
 \theta_2
  \leftrightarrow \Omega R, \Omega R \theta_1, \Omega R \theta_2, \Omega R \theta_1
  \theta_2
\end{equation}
 not intersecting among
themselves. Since this is the most nontrivial setup, this proves
that every D$p$ and O$p$ planes can be mapped to D6/O6 planes, which
are desirable in the $M$-theory dual picture. These planes are
(\ref{z2z2oplane}). It is stressed that both pictures are
equivalent.

For supersymmetric system, we cannot have D-branes other than the
O9/D9 and/or O5/D5 considered above. By $T$-duality, fixing one
orientifold as O9, let us then count the lower dimensional
orientifolds. Consider $T^n/\Z_N$ orbifold \cite{AFIV,BGK}. For
odd $N$, the only possible orientifold is O9 since there is no
even order element in $P$ compatible with lower dimensional
orientifold. For even $N$, the only possibility other than O9 is
O5, because in order not to have a tachyon, the the difference of
dimensions of D-branes should be a multiple of 4. We cannot put
additional O1, if we want four dimensional Lorentz invariance. The
number of O5 is always 1, and it sits on the plane generated by
order 2 element $g^{N/2}$. The same is also applied to $T^n/(\Z_M
\times \Z_N)$ orbifolds, where the number of O5 can be up to 3.
Other systems containing O$p$ and/or O$(p-4)$ can be $T$-dualized
to the above models.

\subsubsection*{An example}

One can see, for example the model considered by Cvetic, Shiu and
Uranga \cite{CSU} is deformable to the one by Berkooz and Leigh
\cite{BL}. Both of them are based on $T^6/(\Z_2\times \Z_2)$
orbifold compactification, defined above. The orbifold actions
determine three $5+1$ dimensional hyperplanes and harbor O5 planes.
With Chan--Paton projector (\ref{CPprojector}), the resulting gauge
group is multiple of $Sp(k)$ with integer $k$, which is embedded in
$SO(32)$, for each O5 planes. It is $SO(32)$ because the product of
the RR charge of O$p$ and the number of fixed points is always 16.
In the maximal group case (without Wilson lines, or when all the
D-branes are on top of one orientifold planes) we have only gauge
group $Sp(8)^4$ which will be the final unification group obtainable
from brane deformations. Nevertheless they are {\em broken
$SO(32)^4$} because the resulting 99 and $5_i 5_i$ spectra in
\cite{BL} can be explained by branching rules ${\bf 496 \to 136} + 3
\cdot {\bf
 120}$. Note that we have
$T$-dual symmetries exchanging O9 with one of the O5$_i$. Therefore
every gauge group arising from D$p$-D$p$ branes can be {\em embedded
into  $SO(32)$, for each orientifold planes}.

Applying $T$-duality in 579 direction we have four O6. By brane
deformations, we can see the spectrum of \cite{CSU} consists of 1/4
cycles only and thus can be deformed to parallel branes on top of
orientifold planes. We can always find the directions in which
$T$-duality maps all O9,D9,O5,D5 to six dimensional objects: O6 and
D6. Therefore we can lift of the model to $M$-theory on $G_2$
manifold picture, where the desirable objects are six dimensional
objects in Type IIA theory, which become geometric objects, i.e.
singularities.

One may note that not every vacua might be connected, since the
deformation of orbifold/orientifold images should be always
deformed together. However it is strongly restricted by anomaly
cancelation of representation as $Spin(32)/\Z_2$ \cite{BLPSSW}.

\section{Discussion}

All the above theory is to be interpreted as type I
compactification, even if we had, in addition to O9, more O5 planes
hence additional gauge groups. It is analogous to the presence of
also additional, twisted closed strings, which were not present in
the untwisted theory. Such extra orientifold sectors are required to
cancel anomalies from twisted strings. Thus depending on the number
of O-planes we may meet more than one $SO(32)$ gauge groups and
their adjoints, which are further broken by projection associated
with orbifold actions. The maximal number is four for $\Z_M \times
\Z_N$ orbifold if $M,N$ are all even. The maximal gauge group is
$SO(32)$ of type I, $U(16)^2$ \cite{GP}, $Sp(8)^4$ \cite{BL} with a
single, two, four O-plane(s), respectively. So all the quarks and
leptons are not necessarily belong to the same adjoint. The upper
limit of the rank 32 constrains realistic models, for example the
racetrack model, making use of high rank group.

A situation is possible where there is no compact dimensions. In
this case, type I and type II theories are indistinguishable since
we can move off all the O$p$-planes $(p<9)$ to infinity, which is as
good as having no O-planes at all. The only exception is the $p=9$
case discussed above, since O9 is space filling and in touch. We
have considered the flat internal space (with singularities), but we
can consider a generic space such as Calabi--Yau. In a curved space,
the supersymmetric condition for intersecting branes is different
and there is no (or not clear) isometry, so that the meaning of
$T$-duality is not clear. It would be an interesting direction to
find whether the tadpole condition still implies general
supersymmetric open string theory to be $T$-dual to type I theory.

The brane recombination is responsible for symmetry breaking and
restoration. In particular, if the local relation of supersymmetry
is satisfied, there is no energy cost on the deformation. This
connects the orbifolded theory to intersecting brane models. During
recombination there is no change of the sum of RR charges of branes.
Since type I string theory is free of gauge and/or gravitational
anomalies and K-theory anomalies, there is no anomalies in the low
energy theory. This should be, since the latter is understood as
spontaneously broken theory of the former. Conceptually, it is not
necessary to have Grand Unification at the intermediate scale.

Although we have a single gauge coupling, above unification scale,
of type I string theory, we might have different gauge coupling due
to tilted brane in the low energy, which is the original property of
DBI action. Such various gauge couplings do not require as many
moduli, but arise from the nonabelian structure the magnetic fluxes
possess. Brane recombination changes the effective gauge couplings
from the unified one, whose dynamics is at the scale of unification.
Gauge couplings can be set be equal by some symmetry, or their
discrepancy may be cured from threshold corrections of magnetic
fluxes. Or, we might not need one meeting point of unification,
running from the low energy. In the coupling unification limit, we
have a desirable value of weak mixing angle $\sin^2 \theta_W= 3/8$
at the unification scale. This is due to embedability of quantum
numbers in $SO(10)$.

\section*{Acknowledgments}

The author is grateful to Ralph Blumenhagen,
Daniel Cremades, Hans-Peter Nilles, Koji Hashimito, Oscar
Loaiza-Brito, Seok Kim,
Hyun-Seok Yang and Piljin Yi for useful discussions.
This work was partially supported by the European Union 6th
Framework Program MRTN-CT-2004-503369 Quest for Unification and
MRTN-CT-2004-005104 ForcesUniverse.


\begin{thebibliography}{99}

%\cite{Berkooz:1996km}
\bibitem{BDL}
M.~Berkooz, M.~R.~Douglas and R.~G.~Leigh,
%``Branes intersecting at angles,''
Nucl.\ Phys.\ B {\bf 480} (1996) 265. [arXiv:hep-th/9606139].
%%CITATION = HEP-TH 9606139;%%

%\cite{Blumenhagen:2005mu}
\bibitem{BCLSU}
See, for example,
  R.~Blumenhagen, M.~Cvetic, P.~Langacker and G.~Shiu,
  %``Toward realistic intersecting D-brane models,''
  Ann.\ Rev.\ Nucl.\ Part.\ Sci.\  {\bf 55} (2005) 71
  [arXiv:hep-th/0502005];
  %%CITATION = HEP-TH 0502005;%%
%\cite{Uranga:2003pz}
%\bibitem{Uranga:2003pz}
  A.~M.~Uranga,
  %``Chiral four-dimensional string compactifications with intersecting
  %D-branes,''
  Class.\ Quant.\ Grav.\  {\bf 20} (2003) S373
  [arXiv:hep-th/0301032],
  %%CITATION = HEP-TH 0301032;%%
and references therein.

%\cite{Georgi:1974yf}
\bibitem{GQW}
  H.~Georgi, H.~R.~Quinn and S.~Weinberg,
  %``Hierarchy Of Interactions In Unified Gauge Theories,''
  Phys.\ Rev.\ Lett.\  {\bf 33}, 451 (1974);
  %%CITATION = PRLTA,33,451;%%
%\cite{Dimopoulos:1981yj}
%\bibitem{Dimopoulos:1981yj}
  S.~Dimopoulos, S.~Raby and F.~Wilczek,
  %``Supersymmetry And The Scale Of Unification,''
  Phys.\ Rev.\ D {\bf 24} (1981) 1681;
  %%CITATION = PHRVA,D24,1681;%%
%\cite{Giunti:1991ta}
%\bibitem{Giunti:1991ta}
  C.~Giunti, C.~W.~Kim and U.~W.~Lee,
  %``Running coupling constants and grand unification models,''
  Mod.\ Phys.\ Lett.\ A {\bf 6} (1991) 1745.
  %%CITATION = MPLAE,A6,1745;%%

%\cite{Blumenhagen:2003jy}
\bibitem{BLS}
R.~Blumenhagen, D.~Lust and S.~Stieberger,
%``Gauge unification in supersymmetric intersecting brane worlds,''
JHEP {\bf 0307} (2003) 36 [arXiv:hep-th/0305146].

%%CITATION = HEP-TH 0305146;%%
\bibitem{CLLL}
M.~Cvetic, P.~Langacker, T.~j.~Li and T.~Liu,
%``D6-brane splitting on type IIA orientifolds,''
Nucl.\ Phys.\ B {\bf 709} (2005) 241. [arXiv:hep-th/0407178].
%%CITATION = HEP-TH 0407178;%%

%\cite{Choi:2006hm}
\bibitem{Choi06}
  K.-S.~Choi,
  %``Unification in intersecting brane models,''
  Phys.\ Rev.\ D {\bf 74}, 066002 (2006)
  [arXiv:hep-th/0603186].
  %%CITATION = HEP-TH 0603186;%%

%\cite{Hashimoto:2003pu}
\bibitem{HT}
K.~Hashimoto and W.~Taylor,
%``Strings between branes,''
JHEP {\bf 0310} (2003) 40. [arXiv:hep-th/0307297].
%%CITATION = HEP-TH 0307297;%%

%\cite{Cremades:2002cs}
\bibitem{CIM02}
D.~Cremades, L.~E.~Ibanez and F.~Marchesano,
%``Intersecting brane models of particle physics and the Higgs mechanism,''
JHEP {\bf 0207} (2002) 022. [arXiv:hep-th/0203160].
%%CITATION = HEP-TH 0203160;%%

%\cite{Erdmenger:2003kn}
\bibitem{EGHK}
J.~Erdmenger, Z.~Guralnik, R.~Helling and I.~Kirsch,
%``A world-volume perspective on the recombination of intersecting branes,''
JHEP {\bf 0404} (2004) 064. [arXiv:hep-th/0309043].
%%CITATION = HEP-TH 0309043;%%

%\cite{Douglas:2004yv}
\bibitem{DZ}
M.~R.~Douglas and C.-g.~Zhou,
%``Chirality change in string theory,''
JHEP {\bf 0406} (2004) 014. [arXiv:hep-th/0403018].
%%CITATION = HEP-TH 0403018;%%

%\cite{Kumar:2006yg}
\bibitem{KW}
J.~Kumar and J.~D.~Wells,
%``Multi-brane recombination and standard model flux vacua,''
arXiv:hep-th/0604203.
%%CITATION = HEP-TH 0604203;%%


%\cite{Choi:2005pk}
\bibitem{CK}
K.-S.~Choi and J.~E.~Kim,
%``Gauge unification via stable brane recombination,''
JHEP {\bf 0511} (2005) 043. [arXiv:hep-th/0508149];
%%CITATION = HEP-TH 0508149;%%
%\cite{Choi:2006ze}
%\bibitem{Choi:2006ze}
K.-S.~Choi,
%``Gauge unification via stable brane recombination,''
AIP Conf.\ Proc.\  {\bf 805} (2006) 346.
%%CITATION = APCPC,805,346;%%

%\cite{Marino:1999af}
\bibitem{MMMS}
M.~Marino, R.~Minasian, G.~W.~Moore and A.~Strominger,
%``Nonlinear instantons from supersymmetric p-branes,''
JHEP {\bf 0001} (2000) 005 [arXiv:hep-th/9911206].
%%CITATION = HEP-TH 9911206;%%

\bibitem{GG}
H.~Georgi and S.~L.~Glashow,
%``Unity Of All Elementary Particle Forces,''
Phys.\ Rev.\ Lett.\  {\bf 32} (1974) 438.
%%CITATION = PRLTA,32,438;%%

%\cite{Hashimoto:1997gm}
\bibitem{AHT}
A.~Hashimoto and W.~Taylor,
%``Fluctuation spectra of tilted and intersecting D-branes from the
%Born-Infeld action,''
Nucl.\ Phys.\ B {\bf 503} (1997) 193. [arXiv:hep-th/9703217].
%%CITATION = HEP-TH 9703217;%%

%\cite{Aharony:1997ju}%\cite{Denef:2000rj}
\bibitem{DST}
F.~Denef, A.~Sevrin and J.~Troost,
%``Non-Abelian Born-Infeld versus string theory,''
Nucl.\ Phys.\ B {\bf 581} (2000) 135 [arXiv:hep-th/0002180].
%%CITATION = HEP-TH 0002180;%%

%\cite{Choi:2004vb}
\bibitem{Choi04} For example,
  K.-S.~Choi,
  %``Spectrum of heterotic string on orbifold,''
  Nucl.\ Phys.\ B {\bf 708} (2005) 194
  [arXiv:hep-th/0405195], and references therein.
  %%CITATION = HEP-TH 0405195;%%

%\cite{'tHooft:1981sz}
\bibitem{tH}
G.~'t Hooft,
%``Some Twisted Selfdual Solutions For The Yang-Mills Equations On A
%Hypertorus,''
Commun.\ Math.\ Phys.\  {\bf 81} (1981) 267.
%%CITATION = CMPHA,81,267;%%

%\cite{vanBaal:1984ar}
\bibitem{vB84}
P.~van Baal,
%``SU(N) Yang-Mills Solutions With Constant Field Strength On T**4,''
Commun.\ Math.\ Phys.\  {\bf 94} (1984) 397.
%%CITATION = CMPHA,94,397;%%

%\cite{Guralnik:1997th}
\bibitem{GR}
Z.~Guralnik and S.~Ramgoolam,
%``From 0-branes to torons,''
Nucl.\ Phys.\ B {\bf 521} (1998) 129. [arXiv:hep-th/9708089].
%%CITATION = HEP-TH 9708089;%%

%\cite{Rabadan:2001mt}
\bibitem{Ra}
  R.~Rabadan,
  %``Branes at angles, torons, stability and supersymmetry,''
  Nucl.\ Phys.\ B {\bf 620} (2002) 152
  [arXiv:hep-th/0107036].
  %%CITATION = HEP-TH 0107036;%%

%\cite{Cremades:2004wa}
\bibitem{CIM}
D.~Cremades, L.~E.~Ibanez and F.~Marchesano,
%``Computing Yukawa couplings from magnetized extra dimensions,''
JHEP {\bf 0405} (2004) 079. [arXiv:hep-th/0404229].
%%CITATION = HEP-TH 0404229;%%

%\cite{Taylor:1997dy}
\bibitem{Ta}
  W.~Taylor,
  %``Lectures on D-branes, gauge theory and M(atrices),''
  arXiv:hep-th/9801182.
  %%CITATION = HEP-TH 9801182;%%

%\cite{Green:1996um}
\bibitem{GGBL}
M.~B.~Green and M.~Gutperle,
%``Light-cone supersymmetry and D-branes,''
Nucl.\ Phys.\ B {\bf 476} (1996) 484 [arXiv:hep-th/9604091];
%%CITATION = HEP-TH 9604091;%%
%\cite{Balasubramanian:1996uc}
%\bibitem{Balasubramanian:1996uc}
V.~Balasubramanian and R.~G.~Leigh,
%``D-branes, moduli and supersymmetry,''
Phys.\ Rev.\ D {\bf 55}, 6415 (1997) [arXiv:hep-th/9611165].
%%CITATION = HEP-TH 961116

%\cite{Sen:1995vr}
\bibitem{Sen1}
  A.~Sen,
  %``A Note on Marginally Stable Bound States in Type II String Theory,''
  Phys.\ Rev.\ D {\bf 54} (1996) 2964
  [arXiv:hep-th/9510229].
  %%CITATION = HEP-TH 9510229;%%5;%%

%\cite{Dasgupta:1997pu}
\bibitem{DM}
K.~Dasgupta and S.~Mukhi,
%``BPS nature of 3-string junctions,''
Phys.\ Lett.\ B {\bf 423} (1998) 261. [arXiv:hep-th/9711094].
%%CITATION = HEP-TH 9711094;%%

%\cite{Angelantonj:1999xf}
\bibitem{AB}
  C.~Angelantonj and R.~Blumenhagen,
  %``Discrete deformations in type I vacua,''
  Phys.\ Lett.\ B {\bf 473} (2000) 86
  [arXiv:hep-th/9911190];
  %%CITATION = HEP-TH 9911190;%%
%\cite{Blumenhagen:2000ea}
%\bibitem{Blumenhagen:2000ea}
  R.~Blumenhagen, B.~Kors and D.~Lust,
  %``Type I strings with F- and B-flux,''
  JHEP {\bf 0102} (2001) 030
  [arXiv:hep-th/0012156].
  %%CITATION = HEP-TH 0012156;%%

%\cite{Blumenhagen:1999ev}
\bibitem{BGK}
  R.~Blumenhagen, L.~Gorlich and B.~Kors,
  %``Supersymmetric 4D orientifolds of type IIA with D6-branes at angles,''
  JHEP {\bf 0001} (2000) 040
  [arXiv:hep-th/9912204].
  %%CITATION = HEP-TH 9912204;%%

%\cite{Witten:1998cd}
\bibitem{Wi98}
  E.~Witten,
  %``D-branes and K-theory,''
  JHEP {\bf 9812} (1998) 019
  [arXiv:hep-th/9810188].
  %%CITATION = HEP-TH 9810188;%%

%\cite{Acharya:2001gy}
\bibitem{KV}
%\cite{Katz:1996xe}
%\bibitem{Katz:1996xe}
S.~Katz and C.~Vafa,
%``Matter from geometry,''
Nucl.\ Phys.\ B {\bf 497} (1997) 146; [arXiv:hep-th/9606086].

%%CITATION = HEP-TH 9606086;%%
\bibitem{AW}
B.~Acharya and E.~Witten,
%``Chiral fermions from manifolds of G(2) holonomy,''
%%CITATION = HEP-TH 0109152;%%
arXiv:hep-th/0109152.

%\cite{Bershadsky:1995qy}
\bibitem{BVS}
  M.~Bershadsky, C.~Vafa and V.~Sadov,
  %``D-Branes and Topological Field Theories,''
  Nucl.\ Phys.\ B {\bf 463} (1996) 420
  [arXiv:hep-th/9511222].
  %%CITATION = HEP-TH 9511222;%%

%\cite{Tseytlin:1997cs}
\bibitem{Ts}
A.~A.~Tseytlin,
%``On non-abelian generalisation of the Born-Infeld action in string  theory,''
Nucl.\ Phys.\ B {\bf 501} (1997) 41
[arXiv:hep-th/9701125].
%%CITATION = HEP-TH 9701125;%%

%\cite{Brecher:1998su}
\bibitem{BP}
D.~Brecher and M.~J.~Perry,
%``Bound states of D-branes and the non-Abelian Born-Infeld action,''
Nucl.\ Phys.\ B {\bf 527} (1998) 121
[arXiv:hep-th/9801127].
%%CITATION = HEP-TH 9801127;%%

%\cite{Bergshoeff:2001dc}
\bibitem{BBRS}
E.~A.~Bergshoeff, A.~Bilal, M.~de Roo and A.~Sevrin,
%``Supersymmetric non-abelian Born-Infeld revisited,''
JHEP {\bf 0107} (2001) 029 [arXiv:hep-th/0105274].
%%CITATION = HEP-TH 0105274;%%

%\cite{Lust:2003ky}
\bibitem{LS}
  D.~Lust and S.~Stieberger,
  %``Gauge threshold corrections in intersecting brane world models,''
  arXiv:hep-th/0302221.
  %%CITATION = HEP-TH 0302221;%%

%\cite{Anastasopoulos:2006da}
\bibitem{ADKS}
P.~Anastasopoulos, T.~P.~T.~Dijkstra, E.~Kiritsis and
A.~N.~Schellekens,
%``Orientifolds, hypercharge embeddings and the standard model,''
arXiv:hep-th/0605226;
%%CITATION = HEP-TH 0605226;%%
%\cite{Antoniadis:2002qm}
%\bibitem{Antoniadis:2002qm}
I.~Antoniadis, E.~Kiritsis, J.~Rizos and T.~N.~Tomaras,
%``D-branes and the standard model,''
Nucl.\ Phys.\ B {\bf 660} (2003) 81 [arXiv:hep-th/0210263].
%%CITATION = HEP-TH 0210263;%%

%\cite{Gimon:1996rq}
\bibitem{GP}
  E.~G.~Gimon and J.~Polchinski,
  %``Consistency Conditions for Orientifolds and D-Manifolds,''
  Phys.\ Rev.\ D {\bf 54} (1996) 1667
  [arXiv:hep-th/9601038].
  %%CITATION = HEP-TH 9601038;%%

%\cite{Aldazabal:1998mr}
\bibitem{AFIV}
G.~Aldazabal, A.~Font, L.~E.~Ibanez and G.~Violero,
%``D = 4, N = 1, type IIB orientifolds,''
Nucl.\ Phys.\ B {\bf 536} (1998) 29. [arXiv:hep-th/9804026].
%%CITATION = HEP-TH 9804026;%%


%\cite{Blumenhagen:2001te}
\bibitem{BKLO}
R.~Blumenhagen, B.~Kors, D.~Lust and T.~Ott,
%``The standard model from stable intersecting brane world orbifolds,''
Nucl.\ Phys.\ B {\bf 616}, 3 (2001) [arXiv:hep-th/0107138].
%%CITATION = HEP-TH 0107138;%%

\bibitem{Heterotic}
For a review, K.-S.~Choi and J.~E.~Kim, ``Quarks and Leptons from
orbifolded superstring,'' Lect. Notes Phys. {\bf 696} (2006), Springer.
%%CITATION = LNPHA,696,1;%%

%\cite{Green:1984sg}
\bibitem{GS}
M.~B.~Green and J.~H.~Schwarz,
%``Anomaly Cancellation In Supersymmetric D=10 Gauge Theory And Superstring
%Theory,''
Phys.\ Lett.\ B {\bf 149} (1984) 117.
%%CITATION = PHLTA,B149,117;%%

\bibitem{BM}
%\cite{Bianchi:2000de}
M.~Bianchi and J.~F.~Morales,
%``Anomalies and tadpoles,''
JHEP {\bf 0003} (2000) 030 [arXiv:hep-th/0002149].
%%CITATION = HEP-TH 0002149;%%

%\cite{Polchinski:1987tu}
\bibitem{PC}
J.~Polchinski and Y.~Cai,
%``Consistency Of Open Superstring Theories,''
Nucl.\ Phys.\ B {\bf 296} (1988) 91.
%%CITATION = NUPHA,B296,91;%%

%\cite{Nilles:1982ik}
\bibitem{Ni}
H.~P.~Nilles,
%``Dynamically Broken Supergravity And The Hierarchy Problem,''
Phys.\ Lett.\ B {\bf 115} (1982) 193.
%%CITATION = PHLTA,B115,193;%%

\bibitem{Mc}
R.~C.~McLean, Commun. Anal. Geom. {\bf 6} (1998) 705.

%\cite{Witten:2000mf}
\bibitem{Wi20}
  E.~Witten,
  %``BPS bound states of D0-D6 and D0-D8 systems in a B-field,''
  JHEP {\bf 0204} (2002) 012
  [arXiv:hep-th/0012054];
  %%CITATION = HEP-TH 0012054;%%
%\cite{Ohta:2001dh}
%\bibitem{Ohta:2001dh}
  K.~Ohta,
  %``Supersymmetric D-brane bound states with B-field and higher dimensional
  %instantons on noncommutative geometry,''
  Phys.\ Rev.\ D {\bf 64} (2001) 046003
  [arXiv:hep-th/0101082].
  %%CITATION = HEP-TH 0101082;%%
See also,
%\cite{Taylor:1997ay}
%\bibitem{Taylor:1997ay}
  W.~I.~Taylor,
  %``Adhering 0-branes to 6-branes and 8-branes,''
  Nucl.\ Phys.\ B {\bf 508} (1997) 122
  [arXiv:hep-th/9705116].
  %%CITATION = HEP-TH 9705116;%%

%\cite{Minasian:1997mm}
\bibitem{MM}
R.~Minasian and G.~W.~Moore,
%``K-theory and Ramond-Ramond charge,''
JHEP {\bf 9711} (1997) 002. [arXiv:hep-th/9710230].
%%CITATION = HEP-TH 9710230;%%

%\cite{Vafa:1994rv}
%\cite{Dabholkar:1996zi}
\bibitem{DP}
  A.~Dabholkar and J.~Park,
  %``An Orientifold of Type-IIB Theory on $K3$,''
  Nucl.\ Phys.\ B {\bf 472}, 207 (1996)
  [arXiv:hep-th/9602030];
  %%CITATION = HEP-TH 9602030;%%
  C.~Vafa and E.~Witten,
  %``On orbifolds with discrete torsion,''
  J.\ Geom.\ Phys.\  {\bf 15} (1995) 189
  [arXiv:hep-th/9409188].
  %%CITATION = HEP-TH 9409188;%%

%\cite{Sagnotti:1987tw}
\bibitem{Sa}
A.~Sagnotti,
``Open Strings And Their Symmetry Groups,'' p. 521 in Proc. of the
1987 Carg\'ese Summer Institute, eds. G. Mack et al. (Pergamon Press,
1988) [arXiv:hep-th/0208020];
%%CITATION = HEP-TH 020802%\cite{Pradisi:1988xd}
%\bibitem{Pradisi:1988xd}
G.~Pradisi and A.~Sagnotti,
%``Open String Orbifolds,''
Phys.\ Lett.\ B {\bf 216} (1989) 59;
%%CITATION = PHLTA,B216,59;%%0;%%
%\cite{Horava:1989vt}
%\bibitem{Horava:1989vt}
P.~Horava,
%``Strings On World Sheet Orbifolds,''
Nucl.\ Phys.\ B {\bf 327}, 461 (1989).
%%CITATION = NUPHA,B327,461;%%

%\cite{Blumenhagen:2000fp}
\bibitem{BGKL}
R.~Blumenhagen, L.~Gorlich, B.~Kors and D.~Lust,
%``Asymmetric orbifolds, noncommutative geometry and type I string vacua,''
Nucl.\ Phys.\ B {\bf 582} (2000) 44
[arXiv:hep-th/0003024].
%%CITATION = HEP-TH 0003024;%%

%\cite{vanBaal:1982ag}
\bibitem{vB82}
  P.~van Baal,
  %``Some Results For SU(N) Gauge Fields On The Hypertorus,''
  Commun.\ Math.\ Phys.\  {\bf 85}, 529 (1982).
  %%CITATION = CMPHA,85,529;%%

%\cite{Guralnik:1997sy}
\bibitem{GR97}
  Z.~Guralnik and S.~Ramgoolam,
  %``Torons and D-brane bound states,''
  Nucl.\ Phys.\ B {\bf 499}, 241 (1997)
  [arXiv:hep-th/9702099].
  %%CITATION = HEP-TH 9702099;%%

%\cite{Witten:1995gx}
\bibitem{Wi95}
E.~Witten,
%``Small Instantons in String Theory,''
Nucl.\ Phys.\ B {\bf 460} (1996) 541. [arXiv:hep-th/9511030].
%%CITATION = HEP-TH 9511030;%%

%\cite{Witten:1994tz}
\bibitem{WiDo}
  E.~Witten,
  %``Sigma Models And The Adhm Construction Of Instantons,''
  J.\ Geom.\ Phys.\  {\bf 15} (1995) 215
  [arXiv:hep-th/9410052];
  %%CITATION = HEP-TH 9410052;%%
%\cite{Douglas:1996uz}
%\bibitem{Douglas:1996uz}
  M.~R.~Douglas,
  %``Gauge Fields and D-branes,''
  J.\ Geom.\ Phys.\  {\bf 28} (1998) 255
  [arXiv:hep-th/9604198].
  %%CITATION = HEP-TH 9604198;%%


%\cite{Cvetic:2001tj}
\bibitem{CSU}
M.~Cvetic, G.~Shiu and A.~M.~Uranga,
%``Three-family supersymmetric standard like models from intersecting  brane
%worlds,''
Phys.\ Rev.\ Lett.\  {\bf 87} (2001) 201801.
[arXiv:hep-th/0107143].
%%CITATION = HEP-TH 0107143;%%

%\cite{Joyce:2001nm}
\bibitem{Jo}
D.~Joyce,
%``Lectures on special Lagrangian geometry,''
arXiv:math.dg/0111111.
%%CITATION = MATH-DG 0111111;%%


%\cite{Berkooz:1996dw}
\bibitem{BL}
M.~Berkooz and R.~G.~Leigh,
%``A D = 4 N = 1 orbifold of type I strings,''
Nucl.\ Phys.\ B {\bf 483} (1997) 187. [arXiv:hep-th/9605049].
%%CITATION = HEP-TH 9605049;%%



%\cite{Dixon:1986jc}
\bibitem{DHVW}
L.~J.~Dixon, J.~A.~Harvey, C.~Vafa and E.~Witten,
%``Strings On Orbifolds. 2,''
Nucl.\ Phys.\ B {\bf 274}, 285 (1986);
%%CITATION = NUPHA,B274,285;%%
%\cite{Dixon:1985jw}
%\bibitem{Dixon:1985jw}
L.~J.~Dixon, J.~A.~Harvey, C.~Vafa and E.~Witten,
%``Strings On Orbifolds,''
Nucl.\ Phys.\ B {\bf 261}, 678 (1985).
%%CITATION = NUPHA,B261,678;%%

\bibitem{EK93} J. Erler and A. Klemm, Commun. Math.
Phys., {\bf 153} (1993) 579; D. G. Markushevich, M. A. Olshanetsky and
A. M. Perelomov, Commun. Math. Phys. 111 (1987) 247-274.



%\cite{Kim:2004pe}
\bibitem{CK38}
J.~E.~Kim,
%``Trinification with sin**2(theta(W)) = 3/8 and seesaw neutrino mass,''
Phys.\ Lett.\ B {\bf 591} (2004) 119
[arXiv:hep-ph/0403196];
%%CITATION = HEP-PH 0403196;%%
%\cite{Choi:2003ag}
%\bibitem{Choi:2003ag}
K.-S.~Choi and J.~E.~Kim,
%``Three family Z(3) orbifold trinification, MSSM and doublet-triplet  splitting
%problem,''
Phys.\ Lett.\ B {\bf 567} (2003) 87
[arXiv:hep-ph/0305002];
%%CITATION = HEP-PH 0305002;%%
%\cite{Buchmuller:2006ik}
%\bibitem{Buchmuller:2006ik}
W.~Buchmuller, K.~Hamaguchi, O.~Lebedev and M.~Ratz,
%``Supersymmetric standard model from the heterotic string. II,''
arXiv:hep-th/0606187;
%%CITATION = HEP-TH 0606187;%%
%\cite{Kim:2006hv}
%\bibitem{Kim:2006hv}
  J.~E.~Kim and B.~Kyae,
  %``String MSSM through flipped SU(5) from Z(12) orbifold,''
  arXiv:hep-th/0608085;
  %%CITATION = HEP-TH 0608085;%%
  arXiv:hep-th/0608086.
  %%CITATION = HEP-TH 0608086;%%

%\cite{Sen:1998sm}
\bibitem{Sen}
A.~Sen,
%``Tachyon condensation on the brane antibrane system,''
JHEP {\bf 9808} (1998) 012
[arXiv:hep-th/9805170].
%%CITATION = HEP-TH 9805170;%%

%\cite{Bianchi:2005yz}
\bibitem{BT}
M.~Bianchi and E.~Trevigne,
%``The open story of the magnetic fluxes,''
JHEP {\bf 0508} (2005) 034 [arXiv:hep-th/0502147].
%%CITATION = HEP-TH 0502147;%%

%\cite{Forste:2000hx}
\bibitem{FHS}
S.~Forste, G.~Honecker and R.~Schreyer,
%``Supersymmetric Z(N) x Z(M) orientifolds in 4D with D-branes at angles,''
Nucl.\ Phys.\ B {\bf 593} (2001) 127
[arXiv:hep-th/0008250].
%%CITATION = HEP-TH 0008250;%%

%\cite{Blumenhagen:2000eb}
\bibitem{BBH}
R.~Blumenhagen, V.~Braun and R.~Helling,
%``Bound states of D(2p)-D0 systems and supersymmetric p-cycles,''
Phys.\ Lett.\ B {\bf 510} (2001) 311 [arXiv:hep-th/0012157];
%%CITATION = HEP-TH 0012157;%%
%\cite{Karch:1998sj}
%\bibitem{Karch:1998sj}
  A.~Karch, D.~Lust and A.~Miemiec,
  %``N = 1 supersymmetric gauge theories and supersymmetric 3-cycles,''
  Nucl.\ Phys.\ B {\bf 553} (1999) 483
  [arXiv:hep-th/9810254].
  %%CITATION = HEP-TH 9810254;%%

%\cite{Witten:1982fp}
\bibitem{global}
E.~Witten,
%``An SU(2) Anomaly,''
Phys.\ Lett.\ B {\bf 117} (1982) 324;
%%CITATION = PHLTA,B117,324;%%
%\cite{Berkooz:1996iz}

%\cite{Freed:1999vc}
\bibitem{kthconst}
D.~S.~Freed and E.~Witten,
%``Anomalies in string theory with D-branes,''
arXiv:hep-th/9907189;
%%CITATION = HEP-TH 9907189;%%
%\cite{Marchesano:2004xz}
%\bibitem{Marchesano:2004xz}
F.~Marchesano and G.~Shiu,
%``Building MSSM flux vacua,''
JHEP {\bf 0411} (2004) 041
[arXiv:hep-th/0409132].
%%CITATION = HEP-TH 0409132;%%
\bibitem{BLPSSW}
M.~Berkooz, R.~G.~Leigh, J.~Polchinski, J.~H.~Schwarz, N.~Seiberg and E.~Witten,
%``Anomalies, Dualities, and Topology of D=6 N=1 Superstring Vacua,''
Nucl.\ Phys.\ B {\bf 475} (1996) 115. [arXiv:hep-th/9605184].
%%CITATION = HEP-TH 9605184;%%

\end{thebibliography}
\end{document}